\documentclass[prd,aps,floats,floatfix,nofootinbib,preprintnumbers]{revtex4-1}
\usepackage{amsmath}
\usepackage{amsfonts}
\usepackage{graphicx}
\usepackage{slashed}
\usepackage{dcolumn}
\usepackage{bm}
\usepackage{amssymb}
\usepackage{latexsym}
\usepackage{color}
\usepackage{url}

\newcommand{\pbrac}[1]{\left( #1 \right)}
\newcommand{\tbrac}[1]{\left[ #1 \right]}
\newcommand{\cbrac}[1]{\left\{ #1 \right\}}

\begin{document}

\title{Strongly First-Order Electroweak Phase Transition \\and Classical Scale Invariance}

\author{Arsham Farzinnia}
\email[]{farzinnia@ibs.re.kr}
\affiliation{Center for Theoretical Physics of the Universe\\Institute for Basic Science (IBS), Daejeon 305-811, Republic of Korea}
\author{Jing Ren}
\email[]{jingren2004@gmail.com}
\affiliation{Department of Physics, University of Toronto\\
Toronto ON Canada M5S1A7}

\preprint{CTPU-14-06}

\date{\today}

\begin{abstract}
In this work, we examine the possibility of realizing a strongly first-order electroweak phase transition within the minimal classically scale invariant extension of the standard model (SM), previously proposed and analyzed as a potential solution to the hierarchy problem. By introducing one complex gauge-singlet scalar and three (weak scale) right-handed Majorana neutrinos, the scenario was successfully capable of achieving a radiative breaking of the electroweak symmetry (by means of the Coleman-Weinberg Mechanism), inducing non-zero masses for the SM neutrinos (via the seesaw mechanism), presenting a pseudoscalar dark matter candidate (protected by the $CP$ symmetry of the potential), and predicting the existence of a second $CP$-even boson (with suppressed couplings to the SM content) in addition to the 125~GeV scalar. In the present treatment, we construct the full finite-temperature one-loop effective potential of the model, including the resummed thermal daisy loops, and demonstrate that finite-temperature effects induce a first-order electroweak phase transition. Requiring the thermally-driven first-order phase transition to be sufficiently strong at the onset of the bubble nucleation (corresponding to nucleation temperatures $T_{N} \sim100$-200~GeV) further constrains the model's parameter space; in particular, an $\mathcal O(0.01)$ fraction of the dark matter in the universe may be simultaneously accommodated with a strongly first-order electroweak phase transition. Moreover, such a phase transition disfavors right-handed Majorana neutrino masses above several hundreds of GeV, confines the pseudoscalar dark matter masses to $\sim 1$-2~TeV, predicts the mass of the second $CP$-even scalar to be $\sim 100$-300~GeV, and requires the mixing angle between the $CP$-even components of the SM doublet and the complex singlet to lie within the range $0.2 \lesssim \sin\omega \lesssim 0.4$. The obtained results are displayed in comprehensive exclusion plots, identifying the viable regions of the parameter space. Many of these predictions lie within the reach of the next LHC run.
\end{abstract}

\maketitle

\section{Introduction}\label{intro}

A viable study of baryogenesis involves investigating the Sakharov's conditions \cite{Sakharov:1967dj}: non-conservation of the baryon number, $C$ and $CP$~symmetry violation, and the loss of thermal equilibrium. Starting from a matter-antimatter symmetric state, processes complying with the first two conditions are capable of generating a net baryon asymmetry. The final condition is, however, necessary in order to prevent the relaxation of such created baryon asymmetry back to zero, due to the inverse `washout' processes. In principle, a first-order phase transition, if sufficiently strong, may facilitate the required departure from thermal equilibrium for the asymmetry-generating processes, and an interesting such candidate is the electroweak phase transition \cite{Kuzmin:1985mm}. Although, within the standard model (SM), the finite-temperature effects can give rise to a first-order electroweak phase transition, the required Higgs boson mass to generate a sufficiently strong first-order transition is much lighter \cite{phi/T=1} than the LEP-II lower limits ($M_{h} > 114.4$~GeV at 95\%~C.L. \cite{Barate:2003sz}). As a consequence, a strongly first-order electroweak phase transition cannot be realized within the ordinary SM framework.

Recently, as a potential solution to the hierarchy problem, a classically scale invariant extension of the SM has been constructed \cite{Farzinnia:2013pga,Farzinnia:2014xia}, by minimally adding one complex gauge-singlet state to the $CP$-symmetric scalar sector.\footnote{Similar minimally-extended scale symmetric scalar potentials, in various contexts, are considered in e.g. \cite{SIother}.} In addition to the 125~GeV scalar, the scenario predicts the existence of a second $CP$-even boson, with a radiatively-generated mass and suppressed couplings to the SM content. The invariance of the potential under the $CP$ symmetry results in the $CP$-odd pseudoscalar degree of freedom to always appear in pairs; hence, providing a WIMP dark matter candidate. Furthermore, the scenario contains three (mass degenerate) flavors of the right-handed Majorana neutrinos, inducing non-zero masses for the SM neutrinos via the see-saw mechanism \cite{seesaw}. Within this framework, all the mass scales (including the electroweak scale) are generated dynamically, by means of the Coleman-Weinberg mechanism \cite{Coleman:1973jx}. The parameter space of the proposed framework has been systematically examined and constrained in \cite{Farzinnia:2013pga,Farzinnia:2014xia}, using various theoretical considerations and the available (collider and dark matter) experimental data.

In the current treatment, we concentrate on investigating the possibility of achieving a strongly first-order electroweak phase transition within the parameter space of the proposed classically scale invariant model. Previous studies of a first-order phase transition for other scenarios with scale symmetry can be found in \cite{SIEWPT,Espinosa:2008kw}. We present the complete expression of the finite-temperature one-loop effective potential, including the contributions of the resummed thermal bosonic daisy loops, and demonstrate that the finite-temperature corrections induce a first-order electroweak phase transition. Requiring such a phase transition to be sufficiently strong---as to prevent a washout of the matter-antimatter asymmetry---heavily constrains the parameter space, rendering the model highly predictive; in particular, the scenario is capable of accommodating an $\mathcal O(0.01)$ fraction of the dark matter in the universe, while simultaneously realizing a strongly first-order electroweak phase transition---a conclusion already established within the similar non-scale symmetric models \cite{DM&EWPT}, and confirmed here for the minimal classically scale invariant scenario. Moreover, the dark matter, right-handed Majorana neutrino, and the second $CP$-even scalar masses are predicted to be confined within specific ranges, along with the value of the mixing angle between the $CP$-even components of the SM doublet and the complex singlet. Many of these predictions lie within the reach of the next LHC run.

The paper is organized as follows: in Section~\ref{Vtree}, we provide a concise review of the scenario, as introduced and analyzed in \cite{Farzinnia:2013pga,Farzinnia:2014xia}. We proceed to discuss the zero-temperature one-loop corrections to the potential in Section~\ref{V1LT0}; whereas, the full finite-temperature one-loop effective potential is constructed in Section~\ref{V1LTfin}. The thermally-driven first-order nature of the electroweak phase transition is explicitely exhibited. The conditions for the start and completion of the first-order electroweak phase transition by means of the bubble nucleation and the strength of the phase transition are elaborated in Section~\ref{1EWPT}. Section~\ref{resl} summarizes the obtained results in comprehensive exclusion plots, for representative values of the model's input parameters, and demonstrates explicitly a simultaneous realization of a strongly first-order phase transition and an $\mathcal O(0.01)$ fraction of the dark matter in the universe. The conclusions are outlined in Section~\ref{concl}.

\section{Minimal Scale Invariant Classical Potential}\label{Vtree}

In the minimal classically scale invariant extension of the standard model (SM), introduced and analyzed in \cite{Farzinnia:2013pga,Farzinnia:2014xia}, a scale invariant potential is constructed by adding one complex gauge-singlet scalar. Additionally, in analogy with the scalar sector of the ordinary SM, the extended potential is postulated to respect the $CP$-symmetry. The most general scalar potential, satisfying these requirements, may be written as
\begin{equation}\label{V0}
V^{(0)}(H,S) = \frac{\lambda_1}{6} \pbrac{H^\dagger H}^2 + \frac{\lambda_2}{6} |S|^4 + \lambda_3 \pbrac{H^\dagger H}|S|^2 + \frac{\lambda_4}{2} \pbrac{H^\dagger H}\pbrac{S^2 + S^{*2}} + \frac{\lambda_5}{12} \pbrac{S^2 + S^{*2}} |S|^2 + \frac{\lambda_6}{12} \pbrac{S^4 + S^{*4}} \ ,
\end{equation}
which contains only real and dimensionless couplings. The $H$ and $S$~fields in \eqref{V0} are, respectively, the SM Higgs doublet and the complex singlet,
\begin{equation}\label{HS}
H= \frac{1}{\sqrt{2}}
\begin{pmatrix} \sqrt{2}\,\pi^+ \\ v_\phi+\phi+i\pi^0 \end{pmatrix} \ , \qquad S =\frac{1}{\sqrt 2} \pbrac{v_\eta + \eta + i\chi} \ ,
\end{equation}
where, only the $CP$-even scalars ($\phi$ and $\eta$) acquire a non-zero vacuum expectation value (VEV).  The VEVs of the scenario are, however, generated dynamically by means of the Coleman-Weinberg mechanism \cite{Coleman:1973jx} at the loop level. The fields $\pi^{0, \pm}$ represent the usual Nambu-Goldstone bosons, eaten by the $Z$ and $W^\pm$ gauge fields; whereas, $\phi$ denotes the SM Higgs boson, with the VEV $v_\phi = 246$~GeV. Interestingly, the $CP$ symmetry of the potential protects the pseudoscalar, $\chi$, from decaying \cite{Farzinnia:2013pga}, rendering it a suitable WIMP dark matter candidate \cite{Farzinnia:2014xia}. Component-wise, the quartic interactions in the potential may be expressed as
\begin{equation}\label{V0quart}
\begin{split}
V^{(0)}_{\text{quartic}} =&\, \frac{1}{24} \tbrac{ \lambda_\phi \phi^4 + \lambda_\eta \eta^4 + \lambda_\chi \chi^4 + \lambda_\phi\pbrac{\pi^0\pi^0 + 2 \pi^+ \pi^-}^2}+ \frac{1}{4}\tbrac{\lambda_m^+ \phi^2 \eta^2 + \lambda_m^- \phi^2 \chi^2 +\lambda_{\eta \chi} \eta^2 \chi^2}
\\
&+ \frac{1}{12} \tbrac{\lambda_{\phi} \phi^2 + 3\lambda_m^+ \eta^2 + 3\lambda_m^- \chi^2}\pbrac{\pi^0\pi^0 + 2 \pi^+ \pi^-}\ ,
\end{split}
\end{equation}
with the corresponding convenient definitions of the quartic couplings
\begin{equation}\label{couprel}
\lambda_\phi \equiv \lambda_{1} \ , \quad \lambda_\eta \equiv \lambda_{2} + \lambda_{5} + \lambda_{6} \ , \quad \lambda_\chi \equiv \lambda_{2} - \lambda_{5} + \lambda_{6} \ , \quad \lambda_{\eta \chi} \equiv \frac{1}{3}\lambda_{2} - \lambda_{6} \ , \quad \lambda_m^\pm \equiv \lambda_{3} \pm \lambda_{4} \ .
\end{equation}

The two $CP$-even scalars $\phi$ and $\eta$, acquiring non-zero VEVs, are mixed due to the `Higgs portal' terms $\lambda_3$ and $\lambda_4$ in \eqref{V0} (or, equivalently, $\lambda_m^+$ in \eqref{couprel}). They may be orthogonally rotated into the corresponding diagonal mass basis, according to
\begin{equation}\label{hs}
\begin{pmatrix} \phi\\ \eta \end{pmatrix}
= \begin{pmatrix} \cos\omega & \sin\omega \\ -\sin\omega & \cos\omega \end{pmatrix} \begin{pmatrix} h \\ \sigma \end{pmatrix} \ , \qquad \cot(2\omega) \equiv \frac{1}{4\lambda_m^+} \tbrac{ (\lambda_\eta-\lambda_m^+) \frac{v_\eta}{v_\phi} - (\lambda_\phi-\lambda_m^+) \frac{v_\phi}{v_\eta} }\ ,
\end{equation}
defining the two physical $CP$-even scalar degrees of freedom, $h$ and $\sigma$. In the current framework, the $h$~boson is identified with the 125~GeV scalar discovered by the LHC \cite{LHCnew}, and the implications of such an identification for the model's parameter space were investigated in \cite{Farzinnia:2013pga}, among other theoretical and experimental considerations. The $\sigma$~boson collider searches were further studied in \cite{Farzinnia:2014xia}.

In addition, the scenario includes three flavors of the gauge-singlet right-handed Majorana neutrinos, in order to account for the non-zero masses of the SM neutrinos by means of the see-saw mechanism \cite{seesaw}. The masses of the former are induced via their Yukawa couplings with the complex singlet. Assuming these Yukawa couplings to be flavor-universal for simplicity (resulting in degenerate masses for the three singlet right-handed neutrinos), one obtains for this sector (omitting the kinetic term)
\begin{equation}\label{LRHN}
\mathcal{L}_{\mathcal N} = - \tbrac{Y^\nu_{ij}\, \bar{L}_{\ell}^{i} \tilde{H} \mathcal{N}^{j} + \text{h.c.}} -\frac{1}{2}y^N \mathcal{I}_{3\times3} \pbrac{S + S^*} \bar{\mathcal{N}}^{i}\mathcal{N}^{i} \ ,
\end{equation}
with $\mathcal{N}_{i} = \mathcal{N}_{i}^{c}$ the 4-component right-handed Majorana neutrino spinors, $L_{\ell}^{i}$ the left-handed lepton doublet, and $\tilde{H} \equiv i \sigma^2 H^*$. The pure gauge-singlet sector is postulated to be $CP$-symmetric; hence, the flavor-universal right-handed neutrino Yukawa coupling, $y^N$, is real. The complex Dirac neutrino Yukawa matrix, $Y^\nu_{ij}$, may be ignored for the rest of the discussion, due to its extremely small entities (of the order of the electron Yukawa coupling) \cite{Farzinnia:2013pga}.

\section{Zero-Temperature One-Loop Effective Potential}\label{V1LT0}

As mentioned, the quantum loops dynamically generate non-zero VEVs for the $CP$-even components of the electroweak doublet and the singlet \eqref{HS} by means of the Coleman-Weinberg mechanism \cite{Coleman:1973jx}. In principle, in order to determine the true vacuum of the system, the full one-loop potential (containing all participating degrees of freedom in the loop) must be minimized. This is, however, a formidable task, and may not always be possible to accomplish analytically.

The minimization of the one-loop potential may, nevertheless, be carried out in a perturbative manner, following the Gildener-Weinberg procedure \cite{Gildener:1976ih}. According to this prescription, one may initially only minimize the tree-level potential \eqref{V0} with respect to the fields $H$ and $S$. Defining the radial combination of the $CP$-even scalars according to $\varphi^{2} = \phi^{2} + \eta^{2}$, the tree-level minimization identifies a flat direction between these two scalar states and their (dynamically-generated) VEVs (see \cite{Farzinnia:2013pga} for the relevant details)
\begin{equation}\label{mincond}
\frac{v_\phi^2}{v_\eta^2} = \frac{-3\lambda_m^+(\Lambda)}{\lambda_\phi(\Lambda)}=\frac{\lambda_\eta(\Lambda)}{-3\lambda_m^+(\Lambda)} \ .
\end{equation}
Since, in the full quantum theory, the couplings run as a function of the renormalization scale, the tree-level minimization of the potential necessarily occurs at a definite energy scale, $\Lambda$. The one-loop contributions will then become particularly important along this flat direction, where they remove the flatness and specify the true physical vacuum of the system.

Along the flat direction \eqref{mincond}, the mixing angle relation in \eqref{hs} yields: $\cot \omega = v_{\eta} / v_{\phi}$, and the radial combination $\varphi$ may be projected along the SM $\phi$ direction
\begin{equation}\label{varphi}
\varphi^{2} (\Lambda) =  \phi^{2} (\Lambda) + \eta^{2} (\Lambda) =\frac{\phi^{2} (\Lambda)}{\sin^{2} \omega} \ .
\end{equation}
The following tree-level masses are, then, obtained for the $h$~scalar, $\chi$~pseudoscalar, and the (degenerate) right-handed Majorana neutrinos \cite{Farzinnia:2013pga}
\begin{equation}\label{masstreemincond}
M_h^2 = \frac{v_\phi^2}{3} \tbrac{\lambda_\phi(\Lambda) -3\lambda_m^+(\Lambda) } \ , \quad M_\chi^2 =  \frac{v_\phi^2}{6\lambda_m^+(\Lambda)} \tbrac{3\lambda_m^+(\Lambda)\lambda_m^-(\Lambda) - \lambda_{\phi}(\Lambda)\lambda_{\eta\chi}(\Lambda)} \ ,\quad M_N = y^N v_\phi \sqrt{\frac{2 \lambda_\phi(\Lambda)}{-3\lambda_m^+(\Lambda)}} \ .
\end{equation}
As mentioned, in the current scenario, the $h$~boson is identified with the discovered 125~GeV scalar at the LHC \cite{LHCnew}; i.e., $M_{h} = 125$~GeV. The electroweak Nambu-Goldstone bosons remain massless to all orders in perturbation theory; whereas, the $\sigma$~scalar (massless at tree-level) obtains a radiatively-generated mass at one-loop (see \eqref{ms}).

It has been shown in \cite{Farzinnia:2013pga} that, along the flat direction, the zero-temperature one-loop effective potential of the $\phi$~field may be expressed as\footnote{Note that the tree-level potential vanishes along the flat direction \eqref{mincond}.}
\begin{equation}\label{V1fin}
V^{(0+1)}_{T=0}(\phi) =\beta\, \phi^4 \tbrac{\log \frac{\phi^2}{v_{\phi}^2}-\frac{1}{2}} \ ,
\end{equation}
where, including the massive $W^{\pm}$ and $Z$ vector bosons, top quark, $h$ scalar, $\chi$ pseudoscalar, and the heavy right-handed neutrinos in the loop, we have
\begin{equation}\label{Bp}
\beta = \frac{1}{64\pi^2 v_\phi^4} \pbrac{M_h^4 + M_\chi^4 + 6M_W^4+3M_Z^4 -12 M_t^4 - 6 M_N^4} \ .
\end{equation}
The numerical coefficients of the masses represent the number of degrees of freedom associated with each particle species. Requiring the zero-temperature one-loop effective potential \eqref{V1fin} to be bounded from below for large values of the $\phi$~field corresponds to satisfying the condition $\beta >0$, which in turn dictates the mass relation
\begin{equation}\label{staboneloop}
M_\chi^4 - 6 M_N^4 > 12 M_t^4 - 6M_W^4 - 3M_Z^4 - M_h^4 \ .
\end{equation}
According to the inequality \eqref{staboneloop}, in this minimal setup, the masses of the (degenerate) right-handed Majorana neutrinos and the pseudoscalar dark matter are related, defining the lower bound of $M_{\chi}$ for a given $M_{N}$.\footnote{We emphasize that the relation \eqref{staboneloop} cannot be satisfied within the pure SM, indicating the failure of the Coleman-Weinberg mechanism.} Furthermore, the one-loop effective potential induces a radiatively-generated mass for the $\sigma$~scalar, which serves as the (pseudo) Nambu-Goldstone boson of the scale symmetry \cite{Farzinnia:2013pga}
\begin{equation}\label{ms}
m_\sigma^2 (\omega, M_{\chi}, M_{N}) = 8 \beta\, v_{\phi}^2 \sin^{2} \omega \ .
\end{equation}
This expression is guaranteed to be positive-definite by the inequality \eqref{staboneloop}.

\begin{figure}
\includegraphics[width=.4\textwidth]{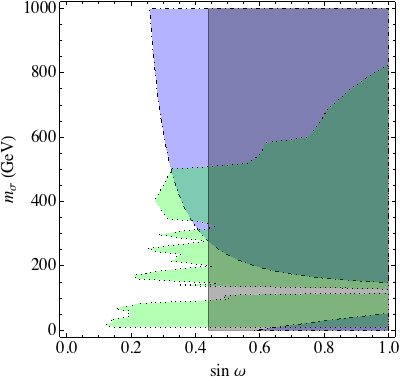}
\caption{The 95\%~C.L. experimental exclusion bounds in the $\sin\omega-m_{\sigma}$~plane. All colored regions are excluded. The LHC direct measurements of the properties of the 125~GeV $h$~scalar (solid vertical line) restrict the mixing angle to $\sin \omega \le 0.44$; whereas, the electroweak precision tests (dot-dashed) and the LEP and LHC Higgs searches (dotted) further reduce the upper bound on the mixing depending on the $\sigma$~boson mass. (Figure taken from \cite{Farzinnia:2014xia})}
\label{LHCexp}
\end{figure}

As such, the current minimal scenario contains five independent inputs, which, without loss of generality, are taken as
\begin{equation}\label{inputs}
\cbrac{\omega, M_\chi, M_N, \lambda_\chi, \lambda_m^-}  \ .
\end{equation}
All the remaining parameters may be determined in terms of the input parameters \eqref{inputs}. In addition, using \eqref{ms}, one may formally replace either of the first three parameters in \eqref{inputs} by $m_{\sigma}$, as an input. For later convenience, we reiterate that the combined analyses in \cite{Farzinnia:2013pga,Farzinnia:2014xia} of the experimental data from the electroweak precision tests, the LHC measurements of the properties of the 125~GeV $h$~scalar, and the LEP and LHC Higgs searches applied to the $\sigma$~boson excluded, at 95\%~C.L., large values of the mixing angle (see Fig.~\ref{LHCexp}).

\section{Finite-Temperature One-Loop Effective Potential}\label{V1LTfin}

Having reviewed the zero-temperature effective potential at one-loop, let us discuss the appropriate contributions at a finite temperature. At this order, the finite-temperature corrections to the one-loop effective potential \eqref{V1fin}, arising from the massive physical states, may be expressed as
\begin{subequations}
\begin{align}
&V^{(1)}_{T}(\phi,T) = \notag \\
& I_{B} (M_{h}(\phi),T) + I_{B} (M_{\chi}(\phi),T) + 6 I_{B} (M_{W}(\phi),T) + 3I_{B} (M_{Z}(\phi),T) -12I_{F} (M_{t}(\phi),T) -6I_{F} (M_{N}(\phi),T) \label{V1TI} \\
& + J_{B}(M_{h}(\phi),\Pi_{h},T) + J_{B}(M_{\chi}(\phi),\Pi_{\chi},T) + 6J_{B}(M_{W}(\phi),\Pi_{W},T) + 3J_{B}(M_{Z}(\phi),\Pi_{Z},T) \label{V1TJ} \ .
\end{align}
\end{subequations}
In \eqref{V1TI}, the contributions from the relevant bosonic and fermionic degrees of freedom are given by \cite{Dolan:1973qd}
\begin{equation}\label{V1TIdef}
I_{B,F} (M_{i}(\phi),T) \equiv \frac{T^{4}}{2\pi^{2}} \int_{0}^{\infty} dx \, x^{2} \log \tbrac{1 \mp e^{-\sqrt{x^{2}+(M_{i}(\phi)/T)^{2}}}} \ ,
\end{equation}
where, the $-(+)$ sign in the integrand corresponds to bosons (fermions), and $M_{i}(\phi)$ is the field-dependent mass of the particle $i$. The field-dependent mass is obtained from the zero-temperature masses, $M_{i}$ (e.g. \eqref{masstreemincond}), by the replacement
\begin{equation}\label{Mphi}
M_{i}(\phi) = M_{i} \times r_{\phi}\ ,  \qquad \quad r_{\phi} \equiv \frac{\phi}{v_{\phi}} \ ,
\end{equation}
with $v_{\phi} = 246$~GeV, the zero-temperature electroweak VEV.

Furthermore, in order to reduce the danger of overestimating the strength of the electroweak phase transition, we include the contributions from the resummed thermal daisy loops for the bosonic Matsubara zero modes in the finite-temperature effective potential \eqref{V1TJ}, which are defined according to \cite{daisy}
\begin{equation}\label{V1TJdef}
J_{B} (M_{i}(\phi),\Pi_{i},T) \equiv \frac{T}{12\pi} \tbrac{M_{i}(\phi)^{3}-\pbrac{M_{i}(\phi)^{2}+\Pi_{i}}^{3/2}} \ .
\end{equation}
Neglecting the small effects associated with the $g'$ coupling, the thermal masses, $\Pi_{i}$, of different bosonic species in \eqref{V1TJdef} are estimated as
\begin{equation}\label{mtherm}
\begin{split}
&\Pi_{W} = \Pi_{Z} = \frac{11}{6} g^{2} T^{2} \ , \qquad \qquad \Pi_{\chi} = \frac{T^{2}}{24}\tbrac{\lambda_{\chi}+3\lambda_{m}^{-} - \lambda_{\chi\chi hh}-\lambda_{\chi\chi\sigma\sigma}} \ , \\
& \Pi_{h} = \frac{T^{2}}{24}\tbrac{ \pbrac{\frac{9}{2}g^{2} + 6 \, y_{t}^{2}}\cos^{2}\omega + 12 \,(y^{N})^{2} \sin^{2}\omega-\lambda_{hhhh} - \lambda_{\chi\chi hh}-\lambda_{\sigma\sigma hh}} \ ,
\end{split}
\end{equation}
with $y_{t}$ the top quark Yukawa coupling, and the relevant scalar quartic couplings, $i\lambda_{ijkl}$, provided in the Appendix~\ref{FR}.

Accordingly, the full finite-temperature effective potential of the current scenario at one-loop is constructed by the sum of \eqref{V1fin}, \eqref{V1TI} and \eqref{V1TJ}
\begin{equation}\label{V01T}
V^{(0+1)}(\phi,T) = V^{(0+1)}_{T=0}(\phi) + V^{(1)}_{T}(\phi,T) - V^{(1)}_{T}(0,T) \ ,
\end{equation}
which is, for convenience, normalized to zero at $\phi = 0$.\footnote{Note that such normalization automatically removes the finite-temperature contributions from the massless states, such as the photons and gluons, the (radiatively generated) $\sigma$~boson (c.f. \eqref{ms}), and the (gauge-dependent) electroweak Nambu-Goldstone bosons. For recent discussions regarding the irrelevance of the Nambu-Goldstone bosons, see e.g. \cite{NGB}.} It is worth noting that the finite-temperature contributions, originating from \eqref{V1TIdef}, depend only on the dark matter mass, $M_{\chi}$, and the right-handed neutrino mass, $M_{N}$, as free parameters. The resummed daisy contributions \eqref{V1TJdef}, on the other hand, involve the entire input set \eqref{inputs}.

The integral in \eqref{V1TIdef} has, in general, no closed-form solutions; nonetheless, it is possible to construct an approximate analytical expression for \eqref{V1TIdef}, accurate within the (sub-)percent level, by smoothly matching its high- and low-temperature limits \cite{anapprox}
\begin{equation}\label{Vanapprox}
\begin{split}
I_{B} &\simeq \Theta \tbrac{x_{B} - \pbrac{\frac{M_{i}(\phi)}{T}}^{2}} U^{\text{high}}_{B}(3) + \Theta \tbrac{\pbrac{\frac{M_{i}(\phi)}{T}}^{2} - x_{B}} \pbrac{U^{\text{low}}(3) - \delta_{B}T^{4}} \ , \\
I_{F} &\simeq -\Theta \tbrac{x_{F} - \pbrac{\frac{M_{i}(\phi)}{T}}^{2}} U^{\text{high}}_{F}(4) - \Theta \tbrac{\pbrac{\frac{M_{i}(\phi)}{T}}^{2} - x_{F}} \pbrac{U^{\text{low}}(3) - \delta_{F}T^{4}} \ .
\end{split}
\end{equation}
In this analytical approximation, $x_{B} = 9.47134$ for bosons and $x_{F} = 5.47281$ for fermions. The corresponding small shifts $\delta_{B} = 3.19310 \times 10^{-4}$ and $\delta_{F} = 4.60156 \times 10^{-4}$ are introduced for a smooth matching of the functions and their derivatives at the point of transition, and $\Theta$ denotes the Heaviside step function. The $n^{\text{th}}$-order low-temperature expansion of \eqref{V1TIdef} has been determined in \cite{Anderson:1991zb}, and reads
\begin{equation}\label{lowT}
U^{\text{low}}(n) = -e^{-M_{i}(\phi)/T} \pbrac{\frac{M_{i}(\phi)\, T}{2\pi}}^{3/2} T \sum_{\ell=0}^{n} \frac{\Gamma(5/2+\ell)}{2^{\ell}\ell !\, \Gamma(5/2-\ell)} \pbrac{\frac{M_{i}(\phi)}{T}}^{-\ell} \ ,
\end{equation}
whereas, its bosonic and fermionic $n^{\text{th}}$-order high-temperature expansions are given in \cite{Arnold:1992rz}
\begin{equation}\label{highT}
\begin{split}
U^{\text{high}}_{B}(n) =&-\frac{\pi^{2} T^{4}}{90} + \frac{M_{i}(\phi)^{2} T^{2}}{24} - \frac{M_{i}(\phi)^{3} T}{12\pi} - \frac{M_{i}(\phi)^{4}}{64\pi^{2}}\tbrac{\log \frac{M_{i}(\phi)^{2}}{T^{2}} - c_{B}} \\
&+ \frac{M_{i}(\phi)^{2} T^{2}}{2} \sum_{\ell=2}^{n} (-1)^{\ell} \,\frac{(2\ell-3)!! \, \zeta(2\ell-1)}{(2\ell)!! \,(\ell+1)} \pbrac{\frac{M_{i}(\phi)}{2\pi T} }^{2\ell} \ , \\
U^{\text{high}}_{F}(n) =&-\frac{7\pi^{2} T^{4}}{720} + \frac{M_{i}(\phi)^{2} T^{2}}{48} + \frac{M_{i}(\phi)^{4}}{64\pi^{2}}\tbrac{\log \frac{M_{i}(\phi)^{2}}{T^{2}} - c_{F}} \\
&- \frac{M_{i}(\phi)^{2} T^{2}}{2} \sum_{\ell=2}^{n}(-1)^{\ell} \,\frac{(2\ell-3)!! \, \zeta(2\ell-1)}{(2\ell)!! \,(\ell+1)} \pbrac{2^{2\ell - 1}-1} \pbrac{\frac{M_{i}(\phi)}{2\pi T} }^{2\ell} \ .
\end{split}
\end{equation}
Here, $c_{B} = 3/2 + 2 \log (4\pi) - 2\gamma_{E} \simeq 5.40762$ and $c_{F} = c_{B} - 2 \log 4 \simeq 2.63503$, and $\gamma_{E}$ denotes the Euler-Mascheroni constant.

\begin{figure}
\includegraphics[width=.55\textwidth]{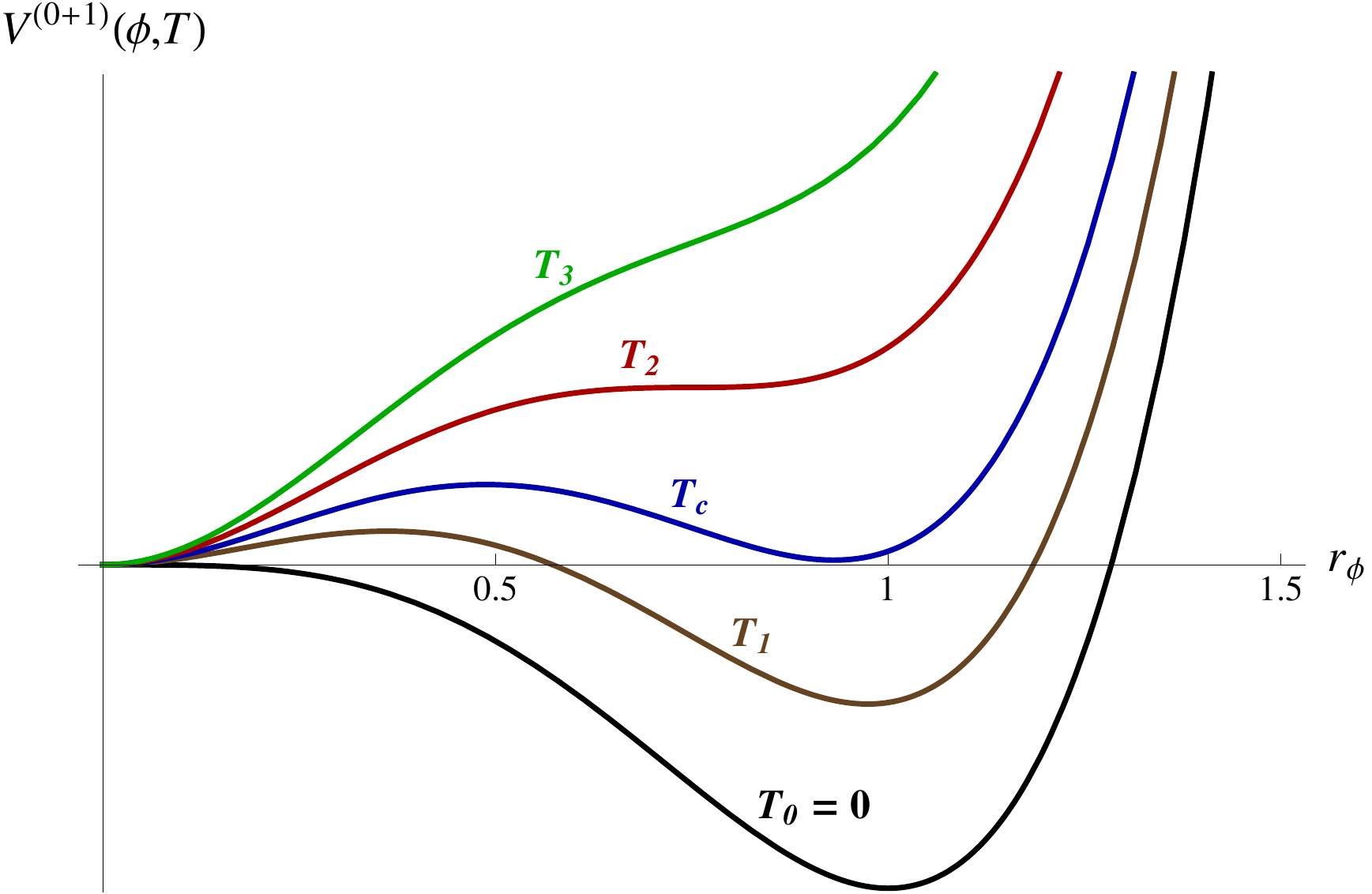}
\caption{Schematic behavior of the (normalized) finite-temperature one-loop effective potential \eqref{V01T} as a function of the (rescaled) field value, $r_{\phi} = \phi / v_{\phi}$ (c.f. \eqref{Mphi}), for various temperatures $T_{0} < T_{1} < T_{c} < T_{2} < T_{3}$. The finite-temperature contributions induce a first-order phase transition. At the temperature $T_{3}$, the potential has a global minimum at $\phi = 0$. As the temperature decreases to $T_{2}$, the potential exhibits an inflection point, where a secondary local minimum starts to develop at $\phi \neq 0$. This minimum at $\phi \neq 0$ becomes degenerate with the one at $\phi = 0$ once the temperature has reached the critical value $T_{c}$, with a barrier separating the two minima. At lower temperatures, $T_{1} < T_{c}$, the minimum at $\phi \neq 0$ becomes global, whereas the minimum at $\phi = 0$ denotes a false vacuum with a diminishing separating barrier. The barrier disappears completely at zero temperature, $T_{0}=0$.}
\label{potT}
\end{figure}

The schematic behavior of the normalized finite-temperature one-loop effective potential \eqref{V01T}, for various relevant temperatures, has been depicted in Fig.~\ref{potT} as a function of the rescaled field, $r_{\phi} = \phi / v_{\phi}$ (c.f. \eqref{Mphi}). At high temperatures in the early universe, the global minimum of the potential is located at the zero field value, $\phi = 0$. As the universe expands and cools, a secondary local minimum starts to appear smoothly with the decreasing temperature, at non-zero values of the field, $\phi \neq 0$, with a barrier separating the two minima. The secondary minimum becomes degenerate with the original minimum at $\phi = 0$ at a critical temperature $T_{c}$, signaling a first-order electroweak phase transition \cite{Quiros:1999jp}, and the height of the barrier reaches its maximum value. As the temperature further decreases, the global minimum of the potential will be located at $\phi \neq 0$, and the barrier shrinks before finally disappearing completely at zero temperature. At this point, the (false) vacuum at $\phi = 0$ vanishes and is replaced by the inflection point of the zero-temperature one-loop effective potential \eqref{V1fin}, characterized by the condition $\dfrac{d^{2} V^{(0+1)}_{T=0}}{d\phi^{2}} \Big |_{\phi =0} = 0$.

It is worth emphasizing that, in the present framework, the development of the barrier separating the two minima---and, therefore, a first-order electroweak phase transition \cite{Quiros:1999jp}---is entirely due to the finite-temperature effects, since the zero-temperature one-loop effective potential \eqref{V1fin} does not possess a barrier (see the $T_{0}=0$ curve in Fig.~\ref{potT}). In the following sections, we explore the constraints on the model's parameter space \eqref{inputs} by demanding such a thermally-driven first-order phase transition to be sufficiently strong as to prevent the washout of a potential baryon asymmetry; an essential ingredient for the successful implementation of a baryogenesis framework at the electroweak scale.

\section{Strongly First-Order Electroweak Phase Transition}\label{1EWPT}

In this section, we investigate the strength of the first-order electroweak phase transition, as induced by the thermal corrections to the one-loop effective potential. In order to prevent the relaxation of a matter-antimatter asymmetry back to zero, the involved asymmetry-creating process must occur out of thermal equilibrium. A first-order electroweak phase transition proceeds by the process of bubble nucleation, where bubbles of the broken phase expand and collide within the universe still in the symmetric phase, until they fill the entire space. As a consequence, during the phase transition, a potential baryon asymmetry-creating process will occur out of thermal equilibrium if the sphaleron rate---responsible for the baryon-asymmetry washout---inside the bubbles is sufficiently suppressed with respect to the outside rate \cite{CKN}. Such a necessary suppression in the sphaleron rate may be achieved if the first-order electroweak phase transition is sufficiently strong, facilitating the required departure from thermal equilibrium for the asymmetry-creating process.

At the critical temperature $T_{c}$ (c.f. Fig.~\ref{potT}), the two minima of the potential are degenerate, and it becomes possible for the system to tunnel, through the separating barrier, from the symmetric phase minimum at the zero field value to the broken phase minimum at the non-zero value of the field. For the scenario at hand, the radius of the expanding bubble, $r_{b}$, turns out to be much larger than the thickness of its wall (see the discussion below \eqref{rT}); therefore, employing the thin-wall approximation formalism is appropriate. Within this approximation, the tunneling rate reads \cite{S3}: $\Gamma(T) = c^{3}_{3} \, T \pbrac{\frac{S_{3}}{2\pi T}}^{3/2} e^{-S_{3}/T}$, with $S_{3}$ the `bounce solution' of the three-dimensional Euclidean action at finite temperature. Along the flat direction of the potential, one obtains
\begin{subequations}
\begin{align}
&S_{3} = \frac{16\pi}{3 \left | V^{(0+1)}\right |^{2}} \, S_{1}^{3} \ , \label{S3} \\
&S_{1} \equiv \int_{0}^{\varphi_{c}} d\varphi \,\sqrt{2V^{(0+1)}(\varphi ,T_{c})} = \frac{1}{\sin \omega}\int_{0}^{\phi_{c}} d\phi \,\sqrt{2V^{(0+1)}(\phi ,T_{c})} \ , \label{S1}
\end{align}
\end{subequations}
where, $V^{(0+1)}(\phi ,T)$ is given by \eqref{V01T}, and $\phi_{c}$ represents the field value (corresponding to the minimum of the potential) at the critical temperature $T_{c}$. In order to derive the final expression on the right-hand side of \eqref{S1}, the relation \eqref{varphi} is employed to project the flat direction along the $\phi$~field. The proportionality constant in the expression for the tunneling rate, $c_{3}$, has the dimension of mass, and may be taken to be either of the quantities $\phi$, $\sqrt {\frac{d^{2} V^{(0+1)}}{d\phi^{2}}}$, $1/r_{b}$, or $T$ \cite{S3}. For the scenario at hand, it turns out that $\phi$, $\sqrt {\frac{d^{2} V^{(0+1)}}{d\phi^{2}}}$, and $T$ constitute the dominant quantities, all of which are approximately of the same order of magnitude.\footnote{In principle, the largest mass scale in the current scenario is the dark matter mass, $M_{\chi}$, which may lie within the TeV region. Nevertheless, since the exponential factor in the tunneling rate expression plays the dominant role, the exact value of the prefactors is of a lesser importance for the analysis.} Hence, we simply set for the tunneling rate
\begin{equation}\label{rate}
\Gamma(T) =  T^{4} \pbrac{\frac{S_{3}}{2\pi T}}^{3/2} e^{-S_{3}/T} \ .
\end{equation}

In order to define the probability for bubble nucleation within a given causal Hubble volume, one needs the temperature-dependent Hubble rate
\begin{equation}\label{HT}
H(T) = \frac{2\pi^{3/2}}{3}\sqrt{\frac{g_{\text{rad}}}{5}} \, \frac{T^{2}}{M_{P}} \ .
\end{equation}
In this expression, $M_{P} = G_{N}^{-1/2} = 1.22\times10^{19}$~GeV denotes the Planck mass, and the number of effective relativistic degrees of freedom is given by $g_{\text{rad}} = 107.75$ in our model \cite{Farzinnia:2014xia}. In this fashion, at a given temperature $T<T_{c}$, the probability for a bubble to be nucleated within a given causal Hubble volume may be expressed as \cite{Anderson:1991zb}
\begin{equation}\label{PT}
P(T) = \int_{T}^{T_{c}} \frac{dT'}{T'} \frac{\Gamma(T')}{H(T')^{4}} \ .
\end{equation}
Although a tunneling through the separating barrier becomes possible at the critical temperature, $T_{c}$, the actual transition does not effectively start until the bubble nucleation probability within a causal volume \eqref{PT} is of order one. Therefore, one may define the nucleation temperature, $T_{N} < T_{c}$, by the condition
\begin{equation}\label{TN}
P(T_{N}) \sim 1 \ .
\end{equation}
Following \cite{Anderson:1991zb}, to the lowest order,\footnote{For algorithms addressing the full numerical computation, see e.g. \cite{numerical}.} one finds from this condition (see also \cite{Espinosa:2008kw})
\begin{equation}\label{S3TN/TN}
\frac{S_{3}(T_{N})}{T_{N}} \sim \log \frac{T_{N}^{4}}{H(T_{N})^{4}} \sim 142 - \log \frac{T_{N}^{4}}{v_{\phi}^{4}} \ ,
\end{equation}
with the temperature-dependent Hubble rate defined in \eqref{HT}. Using the definition \eqref{S3}, one may subsequently extract the nucleation temperature numerically for a given choice of the input parameters, and obtain the corresponding field value $\phi_{N}$, where the potential assumes a global minimum at this temperature. In this fashion, a sufficiently strong first-order electroweak phase transition, necessary for preserving a matter-antimatter asymmetry, may be characterized by the (perturbative) condition\footnote{Technically, this condition (or instead $\phi_{c}/T_{c} \gtrsim 1$, as sometimes employed in the literature) is appropriate for the case of the SM \cite{phi/T=1}. Nevertheless, since the gauge field configurations, rather than the scalar configurations, matter more importantly in the sphaleron energy \cite{Klinkhamer:1984di}, the (perturbative) requirement \eqref{strongEWPTTN} is expected to remain a good approximation within the current framework (see \cite{Fuyuto:2014yia} for an improved condition in the real singlet-extended SM case). Furthermore, non-perturbative effects are expected to enhance the (perturbatively analyzed) strength of the first-order phase transition \cite{nonpert}. For discussions addressing the potential gauge invariance issues, see \cite{nonGI}.}
\begin{equation}\label{strongEWPTTN}
\frac{\phi_{N}}{T_{N}} \gtrsim 1 \ .
\end{equation}

Furthermore, the phase transition is completed once the expanding bubbles of the broken phase fill the entire space. As the bubbles cannot expand faster than the speed of light, it takes some time for them to collide and fill the universe. The temperature, $T_{f}$, at the end of the phase transition is, therefore, lower than the nucleation temperature, $T_{f} < T_{N} < T_{c}$. The fraction of the universe covered by the bubbles, at a temperature $T<T_{c}$, may be defined by the expression
\begin{equation}\label{fT}
f(T) = \frac{4\pi}{3} \int_{T}^{T_{c}} \frac{dT'}{T'} \frac{\Gamma(T')}{H(T')^{4}} \, w^{3} \pbrac{1-\frac{T}{T'}}^{3} \ ,
\end{equation}
with $w$ the velocity of the expanding bubble wall. An upper bound for the temperature $T_{f}$ may, then, be found by assuming $w \sim 1$, which allows for (initial) overlaps between the bubbles to be ignored, and by demanding the condition
\begin{equation}\label{Tf}
f(T_{f}) \sim 1 \ .
\end{equation}
In a manner analogous to $T_{N}$, one may attempt to estimate the quantity $S_{3}(T_{f})/T_{f}$ to the lowest order from the condition \eqref{Tf}, and utilize, once more, the definition \eqref{S3} to extract $T_{f}$ numerically, together with the corresponding field value, $\phi_{f}$, at the global minimum of the potential (see also \cite{Huber:2007vva}).

For the purpose of illustration, however, we confine our further analyses to $T_{N}$; in other words, we demand the electroweak phase transition to be strongly first-order at the onset of the nucleation (condition \eqref{strongEWPTTN}), rather than when the bubbles have filled the entire space ($\phi_{f}/T_{f} \gtrsim 1$). We have verified that both choices lead to practically similar results for the exclusion bounds on the parameter space, and the main conclusions remain unaltered.

\section{Results and Discussions}\label{resl}

Having developed the relevant formal framework in the previous sections, we continue to discuss the consequences of realizing a strongly first-order electroweak phase transition for the model's parameter space. A glance at the definition \eqref{S3} and \eqref{S1} reveals that the nucleation temperature, $T_{N}$ (c.f \eqref{S3TN/TN}), and the corresponding field value, $\phi_{N}$, are related to the input parameter set \eqref{inputs}. Demanding the electroweak phase transition to be strongly first-order at the onset of the bubble nucleation corresponds to requiring the condition \eqref{strongEWPTTN} being satisfied, which, in turn, imposes constraints on the free parameter space.

In the previous analyses of the minimal scenario at hand in \cite{Farzinnia:2013pga,Farzinnia:2014xia}, extensive exclusion plots were presented, illustrating the viable region of its parameter space, by taking into account various theoretical and experimental constraints. The considered exclusion bounds arose from analyzing the perturbative unitarity, stability of the one-loop effective potential \eqref{staboneloop}, electroweak precision tests, direct measurements of the 125~GeV $h$~boson, LEP and LHC Higgs searches, as well as the dark matter direct detection data obtained by the LUX experiment \cite{LUX2013} and the dark matter relic abundance as reported by the Planck collaboration \cite{Ade:2013zuv}. It should be noted that in analyzing the dark matter constraints in \cite{Farzinnia:2014xia}, it was assumed that the $\chi$~pseudoscalar WIMP candidate of the scenario constitutes the sole or dominant component of the dark matter in the universe. It was demonstrated that the experimental bounds from the electroweak precision tests and the collider searches excluded large values of the mixing angle (see Fig.~\ref{LHCexp}), and was concluded that the pseudoscalar-dominant dark matter relic density could be comfortably accommodated within the viable region of the parameter space.

\begin{figure}
\includegraphics[width=.45\textwidth]{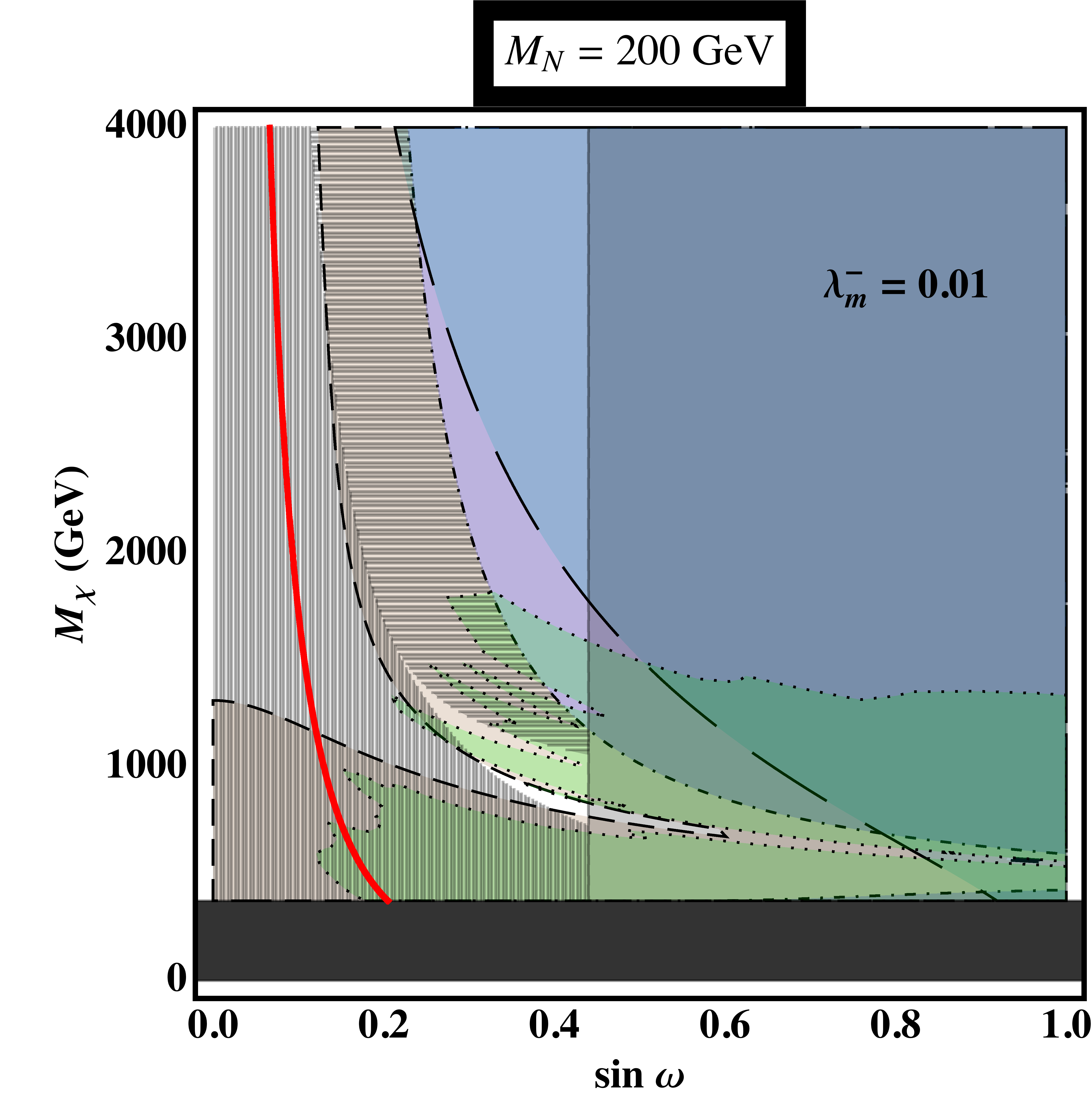}
\includegraphics[width=.45\textwidth]{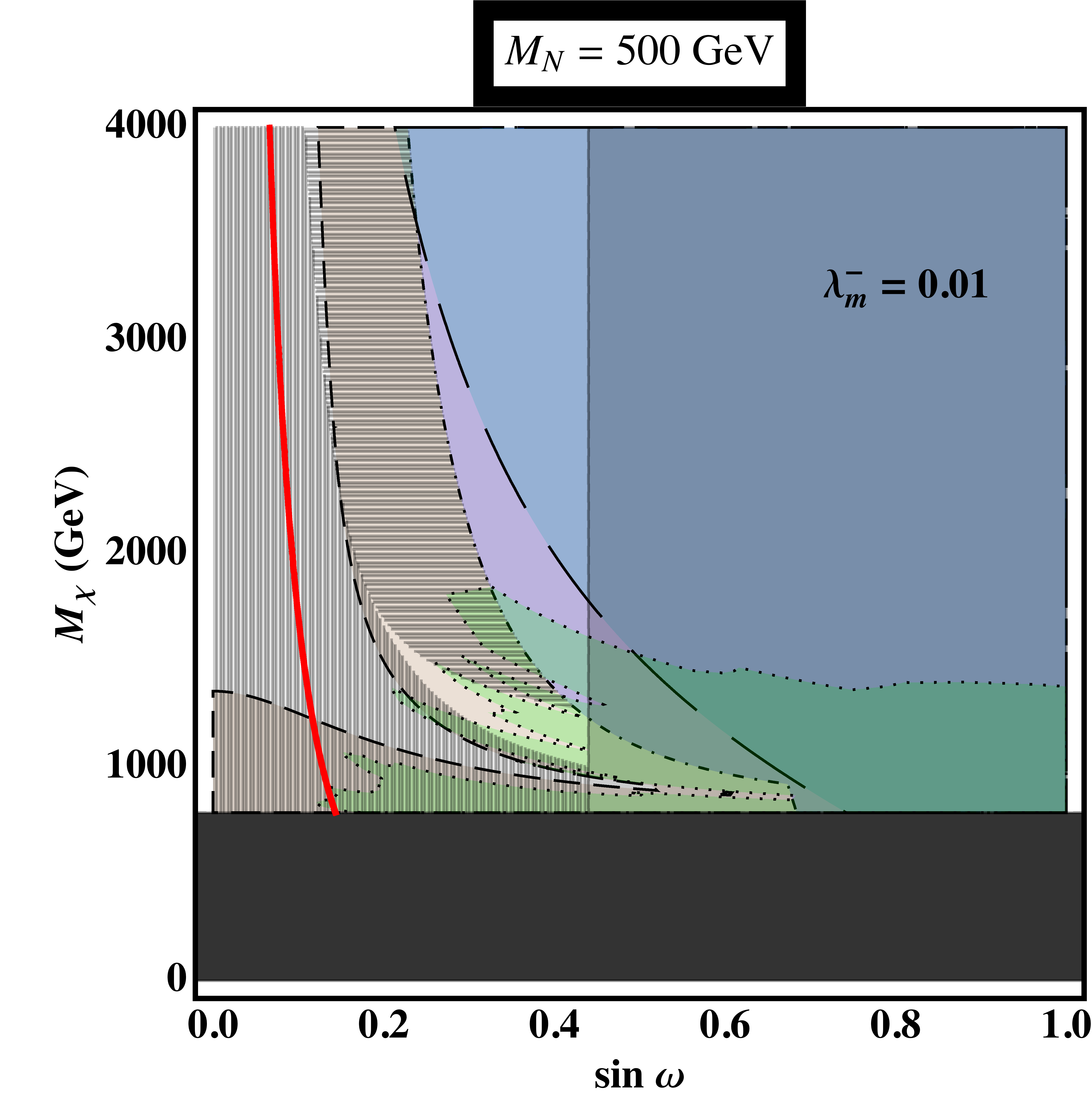}
\caption{Formal and experimental constraints in $\sin\omega - M_{\chi}$ plane, for $\lambda_{\chi}=0.1$ and representative values of $\lambda_{m}^{-}$ and $M_{N}$. All colored regions are excluded. The formal bounds arise due to perturbative unitarity (long-dashed) and the stability of the one-loop effective potential (solid black bar). The 95\%~C.L. experimental bounds arise from the electroweak precision tests (dot-dashed), direct measurements of the properties of the 125~GeV $h$~boson (solid vertical line), and the LEP and LHC Higgs searches (dotted). The thick red band corresponds to the WIMP relic abundance within the $1\sigma$ uncertainty, quoted by the Planck collaboration, assuming the $\chi$~pseudoscalar to constitute an $\mathcal O(1)$~fraction of the dark matter in the universe. The constraints by the LUX dark matter direct detection experiment (short-dashed) at 90\%~C.L. also rest on this assumption. Requiring the framework to accommodate a strongly first-order phase transition at the onset of the bubble nucleation \eqref{strongEWPTTN} further excludes the horizontally-shaded region; whereas, no solution for $T_{N}$ \eqref{TN} exists in the vertically-shaded region. The plots illustrate the tension, within the minimal scenario, between realizing a strongly first-order electroweak transition and the constraints from a pseudoscalar-dominant dark matter component, leaving no viable parameter space.}
\label{MXomSmallruledout}
\end{figure}

\begin{figure}
\includegraphics[width=.43\textwidth]{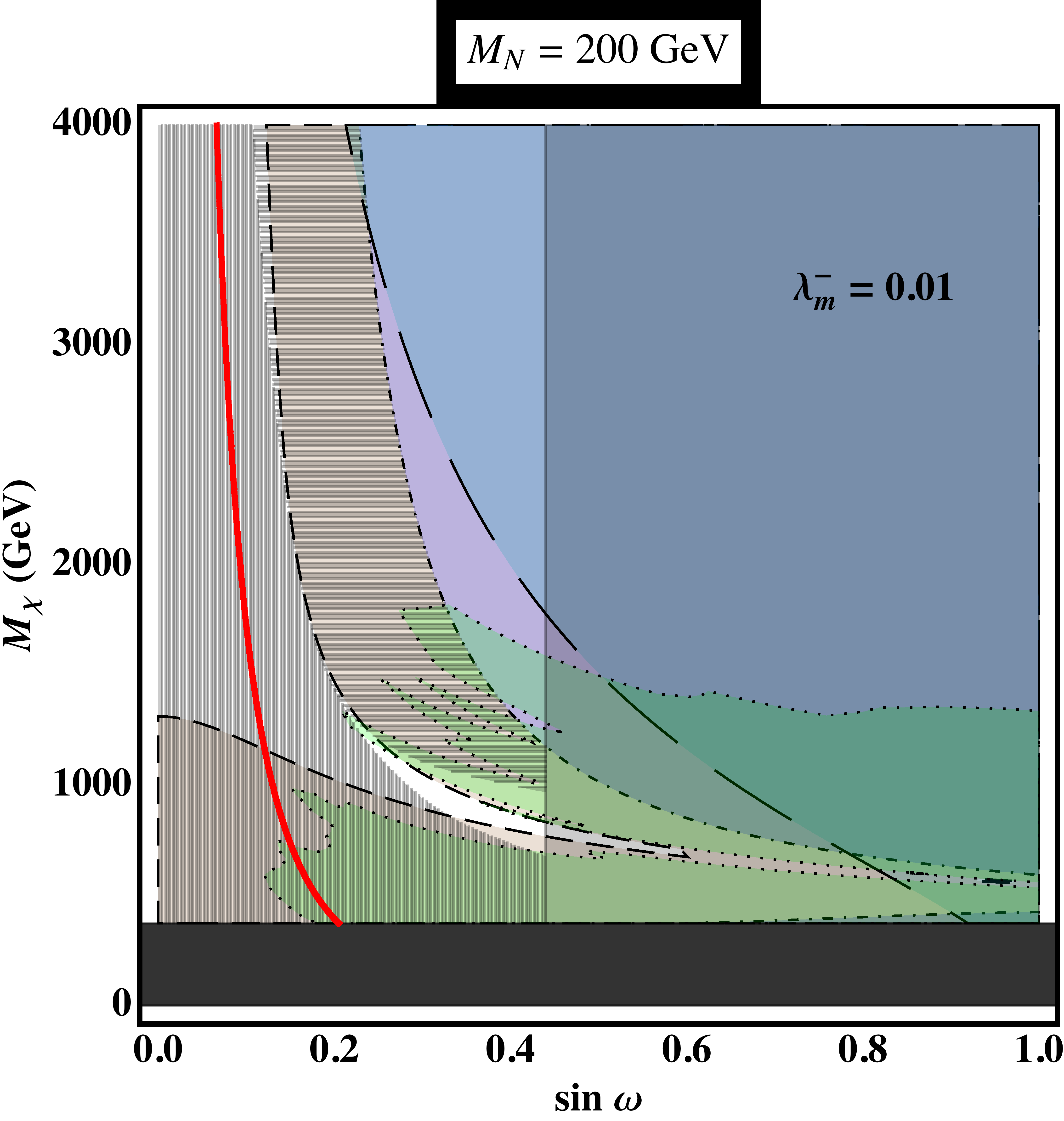}
\includegraphics[width=.43\textwidth]{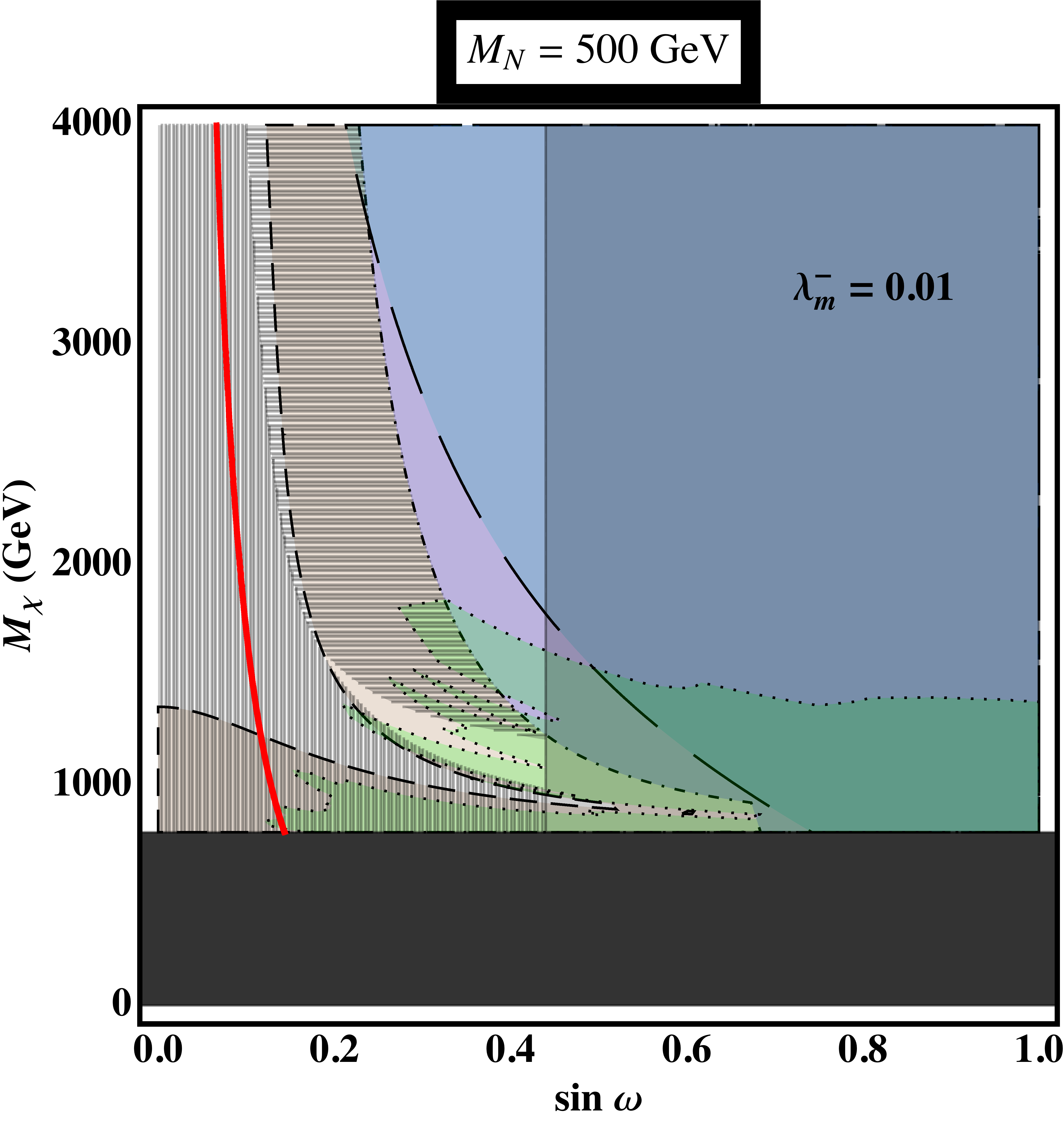}
\includegraphics[width=.43\textwidth]{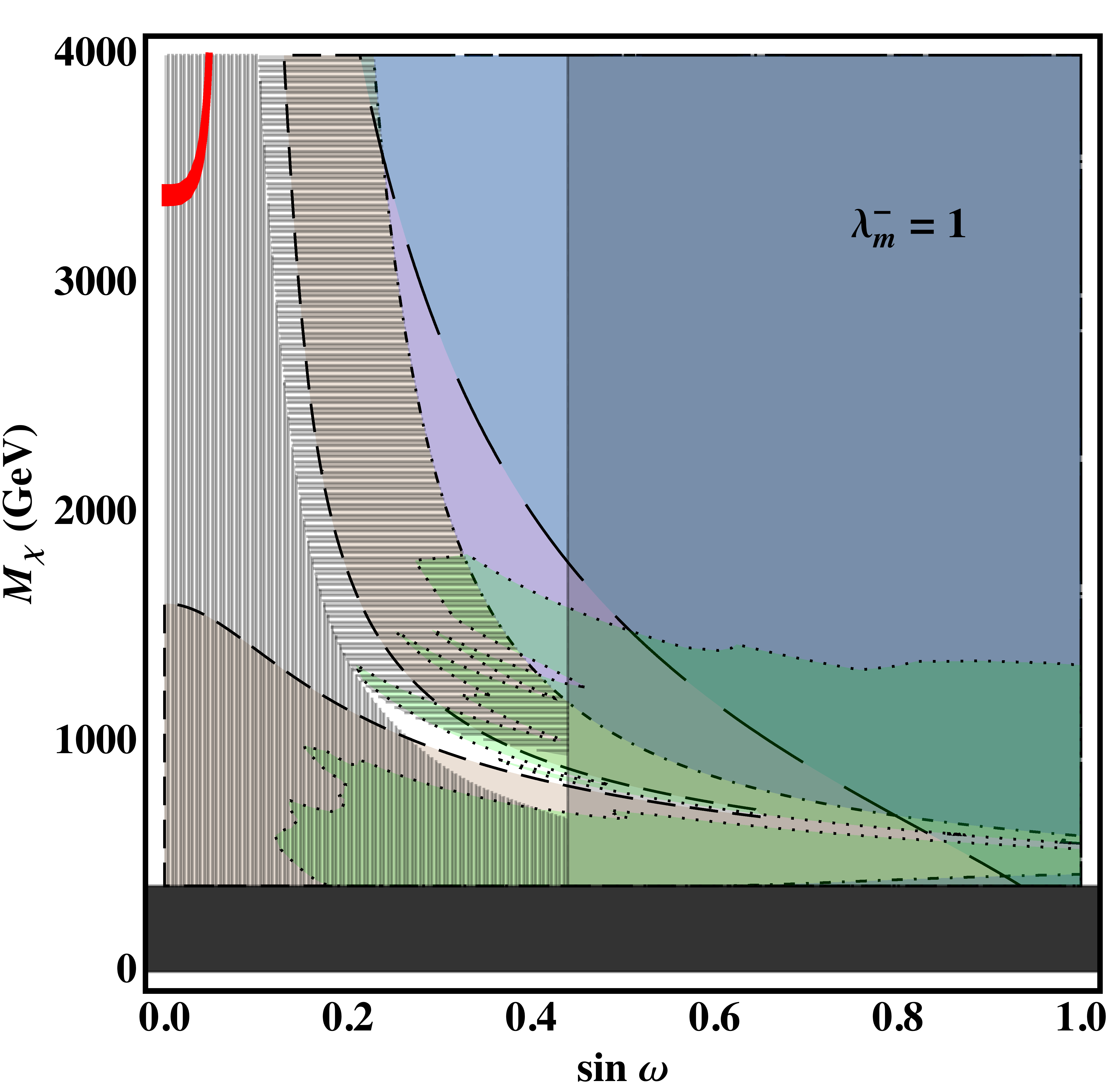}
\includegraphics[width=.43\textwidth]{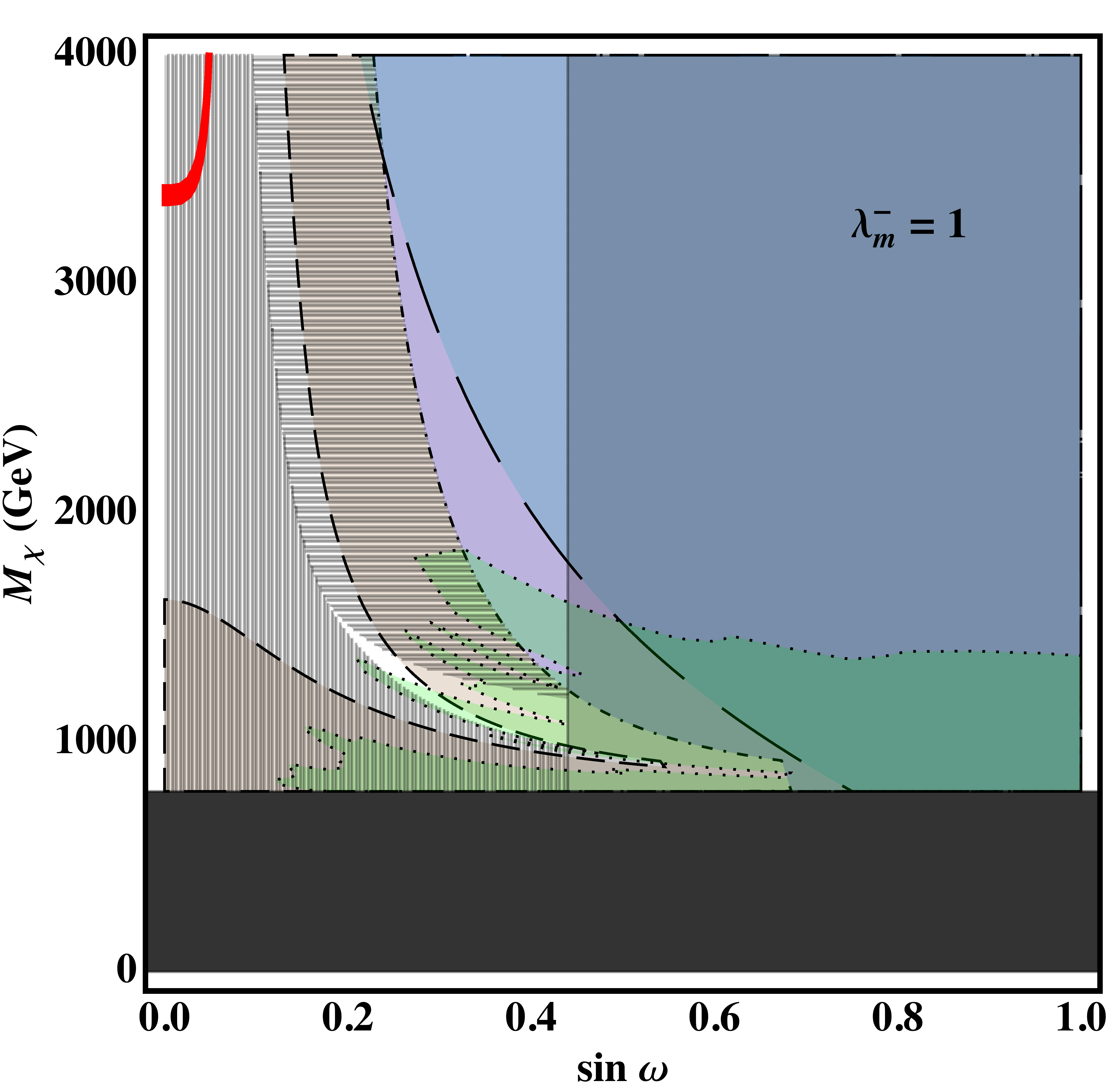}
\includegraphics[width=.45\textwidth]{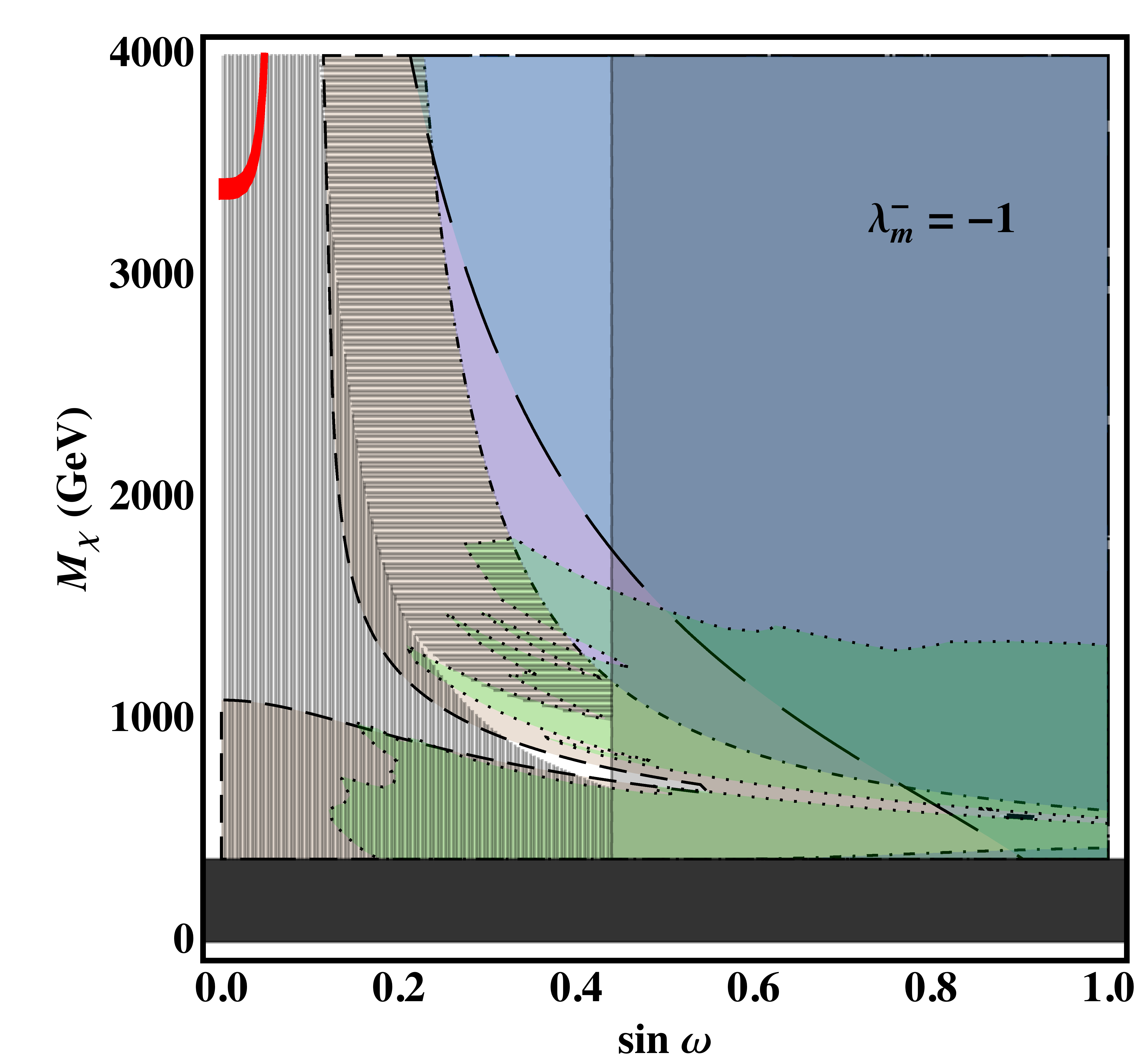}
\includegraphics[width=.43\textwidth]{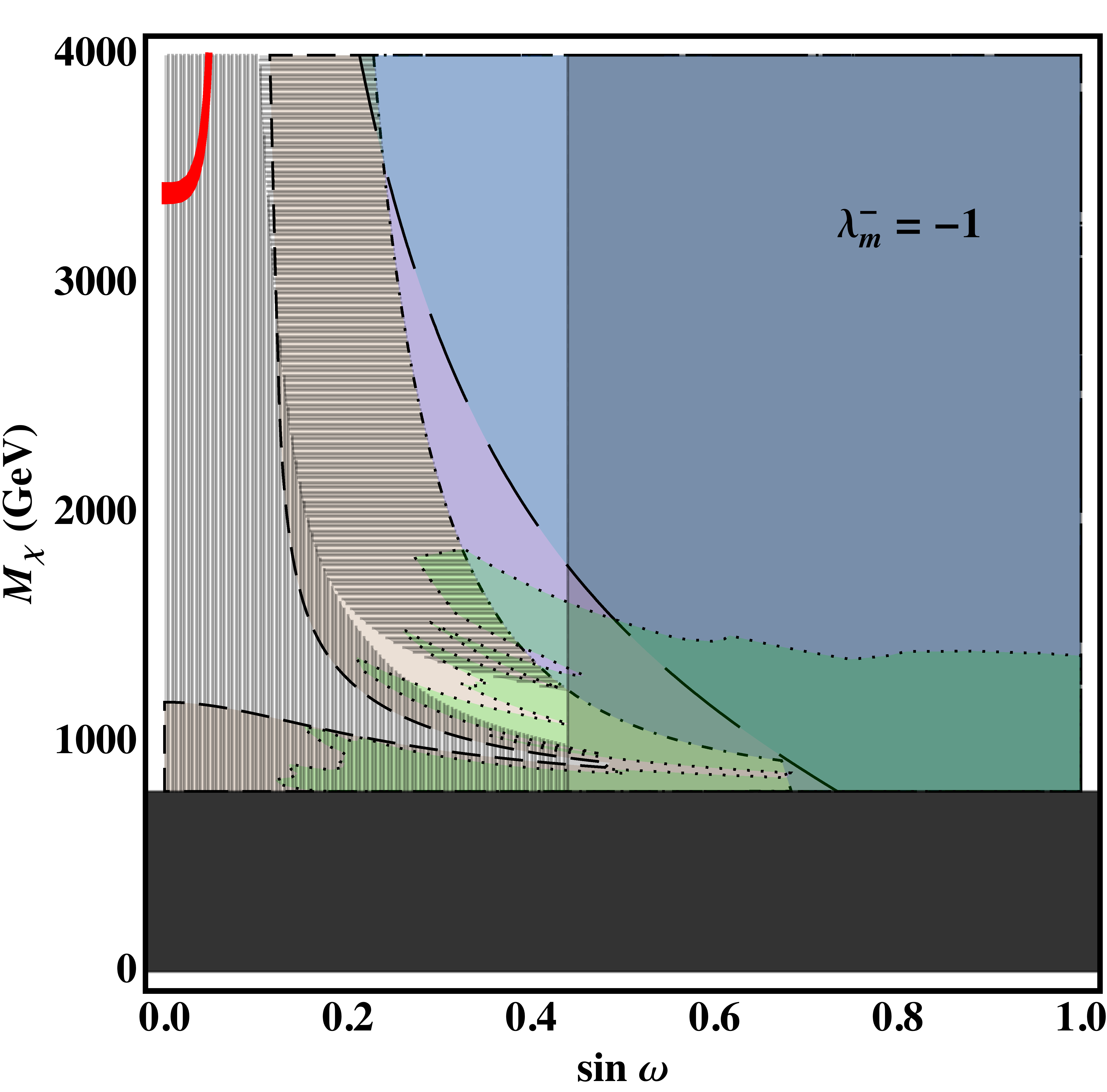}
\caption{Formal and experimental constraints in $\sin\omega - M_{\chi}$ plane, for $\lambda_{\chi}=4\pi$ and representative values of $\lambda_{m}^{-}$ (rows) and $M_{N}$ (columns). All colored regions are excluded. As in Fig.~\ref{MXomSmallruledout}, a simultaneous realization of a strongly first-order electroweak phase transition along with accommodating the entire dark matter in the universe cannot be achieved within the considered minimal framework. (See the caption of Fig.~\ref{MXomSmallruledout} for the details of the plots)}
\label{MXomruledout}
\end{figure}

As mentioned before, demanding the framework to additionally realize a strongly first-order electroweak phase transition (at the onset of the bubble nucleation \eqref{TN}) corresponds to satisfying the condition \eqref{strongEWPTTN}. This requirement imposes considerable restrictions on the previously available parameter space, as demonstrated in Figs.~\ref{MXomSmallruledout}~and~\ref{MXomruledout}, for representative values of $\lambda_{\chi} = 0.1$ and $\lambda_{\chi} = 4\pi$, respectively. In these figures the dark matter mass, $M_{\chi}$, is plotted versus the mixing angle, $\sin \omega$, for various benchmark choices of $\lambda_{m}^{-}$ and $M_{N}$, exhibiting the phase transition constraints (horizontally- and vertically-shaded regions) superimposed on the combined aforementioned theoretical and experimental exclusion bounds \cite{Farzinnia:2013pga,Farzinnia:2014xia}. In the vertically-shaded region, no solution for the nucleation temperature, $T_{N}$, exists; i.e., the condition \eqref{TN} cannot be fulfilled, and the bubble nucleation never starts. This is attributed to the fact that, within this region, $S_{1}$ \eqref{S1} (and, hence, $S_{3}$ \eqref{S3}) is quite large, leading to a small tunneling rate \eqref{rate}. Sufficiently large values of $M_{\chi}$ (for a given $\sin\omega$), however, increase the tunneling rate, and allow for viable solutions of $T_{N}$ to appear. The horizontally-shaded region, on the other hand, excludes the parameter space where such a $T_{N}$ solution does exist, but the corresponding first-order phase transition is too weak to prevent an asymmetry washout; i.e., the condition \eqref{strongEWPTTN} cannot be satisfied in this regime. The two regimes asymptotically approach one another for smaller values of the mixing angle, leaving only a narrow `wedge' of the parameter space viable between them for $0.2 \lesssim \sin\omega \lesssim 0.4$.\footnote{As mentioned, the combined experimental bounds set the upper bound of the mixing angle, excluding larger values (see Fig.~\ref{LHCexp}). For this reason, we do not extend the displayed phase transition constraints (horizontally- and vertically-shaded regimes) beyond the mixing angle values already excluded by the electroweak precision tests and the LHC direct measurements of the 125 GeV $h$~boson.} Within this wedge region, one finds for the nucleation temperature, $T_{N} \sim100$-200~GeV. The wedge region is, however, practically excluded by other considerations---most prominently by the LUX dark matter direct detection bounds---and also cannot accommodate the correct WIMP relic density.


It is evident, from the exclusion plots in Figs.~\ref{MXomSmallruledout}~and~\ref{MXomruledout}, that a strongly first-order electroweak phase transition cannot be accommodated in most of the otherwise unrestricted regions of the parameter space. Most notably, it is severely in tension with the explored dark matter constraints in \cite{Farzinnia:2014xia} for practically all choices of the input parameters, and in particular, rules out the observational value of the thermal relic density within the model. As the depicted dark matter constraints rest on the presumption that the $\chi$~pseudoscalar WIMP comprises the entire dark matter in the universe, one concludes that the current minimal scenario cannot, simultaneously, accommodate a strongly first-order electroweak phase transition along with an $\mathcal O(1)$~fraction of the pseudoscalar dark matter in the universe.\footnote{This conclusion is in accordance with similar findings in the context of the ($Z_{2}$-symmetric) non-scale invariant Higgs portal scenarios \cite{DM&EWPT}.}

Furthermore, it is worth noting that the phase transition exclusion bounds (horizontally and vertically-shaded regimes) are weakly dependent on the precise values of the $\lambda_{m}^{-}$ and $\lambda_{\chi}$ input parameters. Such a dependence finds its origin in the resummed thermal daisy loops \eqref{V1TJdef}, which represent relatively moderate corrections to the dominant contributions \eqref{V1TIdef} within the finite-temperature one-loop effective potential. In contrast, a strongly first-order phase transition disfavors right-handed Majorana neutrinos heavier than several hundreds of GeV, and the viable wedge between the two regions shrinks rapidly close to $M_{N} \sim 1$~TeV.

Requiring, nonetheless, for the minimal scenario to realize a strongly first-order electroweak phase transition, one may investigate what fraction of the dark matter in the universe may be compatibly attributed to the $\chi$~pseudoscalar WIMP. To this end, we define the following relations for the $\chi$~relic abundance and its direct detection cross section
\begin{equation}\label{fDM}
\Omega_{\chi} h^{2} = f_{\text{DM}} \times \Omega_{C} h^{2} \ , \qquad \sigma^{\text{SI}}_{\text{LUX}} = f_{\text{DM}} \times \sigma^{\text{SI}}_{N\chi \to N\chi} \ ,
\end{equation}
where, $f_{\text{DM}} \le1$ represents the fraction of the dark matter composed of the $\chi$~WIMP, $\Omega_{C} h^{2} = 0.1199 \pm 0.0027$ is the relic density value reported by the Planck collaboration \cite{Ade:2013zuv}, and $\sigma^{\text{SI}}_{\text{LUX}}$ denotes the LUX \cite{LUX2013} upper limit on the spin-independent cross section at 90\%~C.L. for a given dark matter mass.\footnote{Accordingly, in the analysis performed in \cite{Farzinnia:2014xia}, $f_{\text{DM}}$ was taken equal to 1.}

\begin{figure}
\includegraphics[width=.43\textwidth]{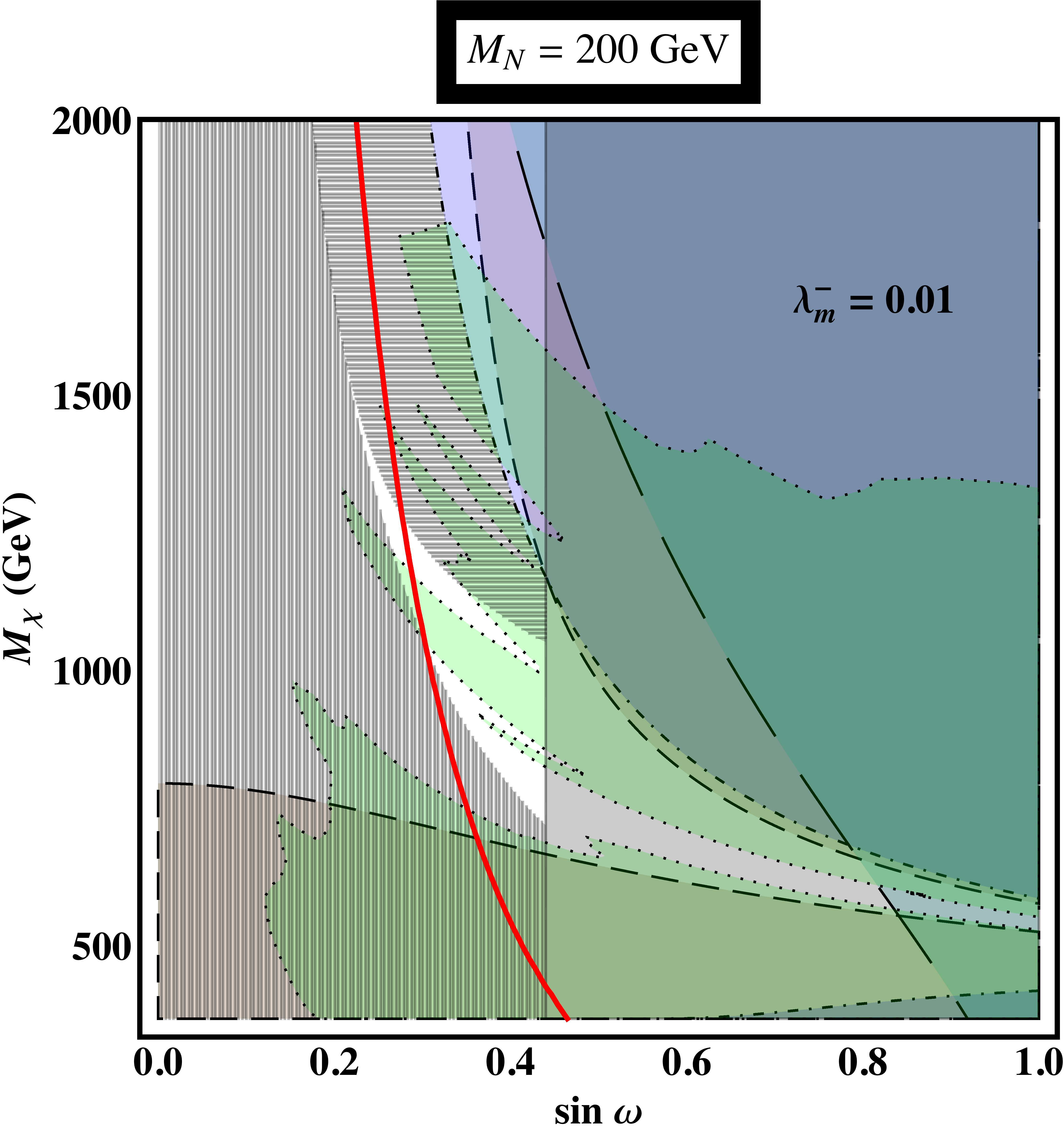}
\includegraphics[width=.45\textwidth]{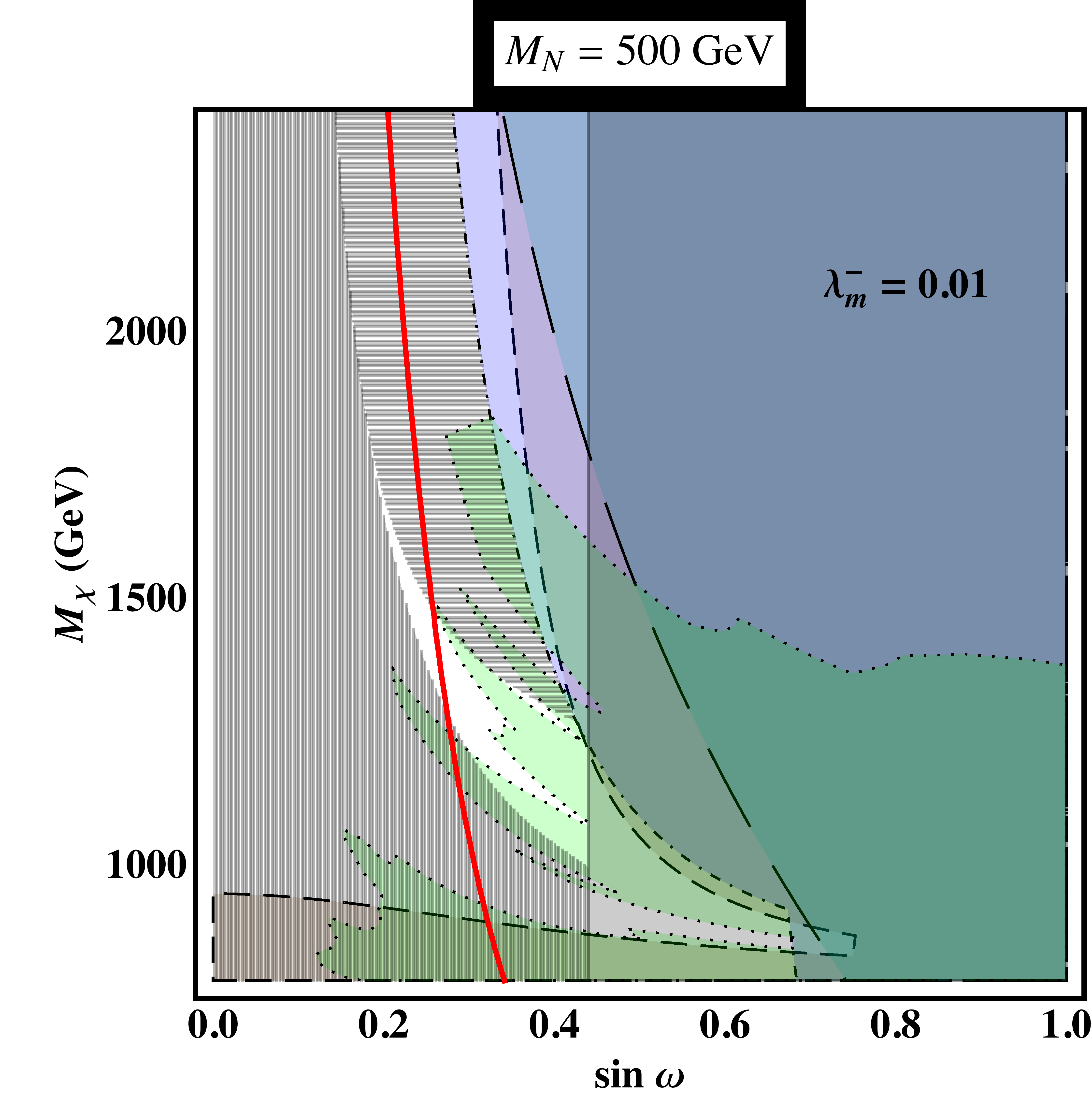}
\caption{Formal and experimental constraints in the enlarged $\sin\omega - M_{\chi}$ plane, for $\lambda_{\chi}=0.1$ and representative values of $\lambda_{m}^{-}$ and $M_{N}$. All colored regions are excluded, with only the uncolored parts of the (center) wedge region unconstrained. The contribution of the $\chi$~pseudoscalar to the dark matter in the universe is chosen to be 3\%, $f_{\text{DM}} = 0.03$, mitigating the tension between the dark matter and strongly first-order phase transition constraints. (See the caption of Fig.~\ref{MXomSmallruledout} for the details of the plots)}
\label{MXomSmall}
\end{figure}

\begin{figure}
\includegraphics[width=.44\textwidth]{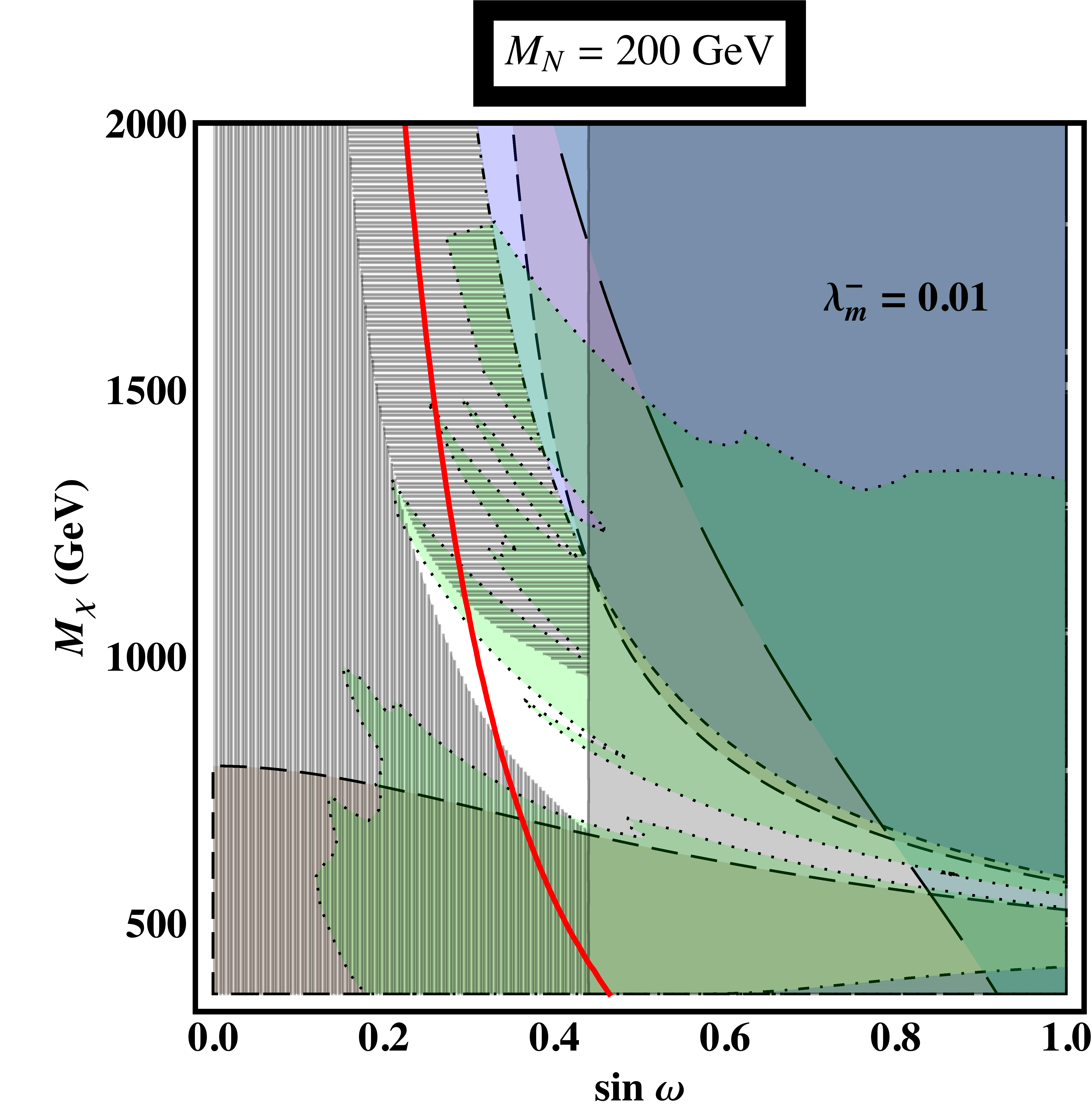}
\includegraphics[width=.44\textwidth]{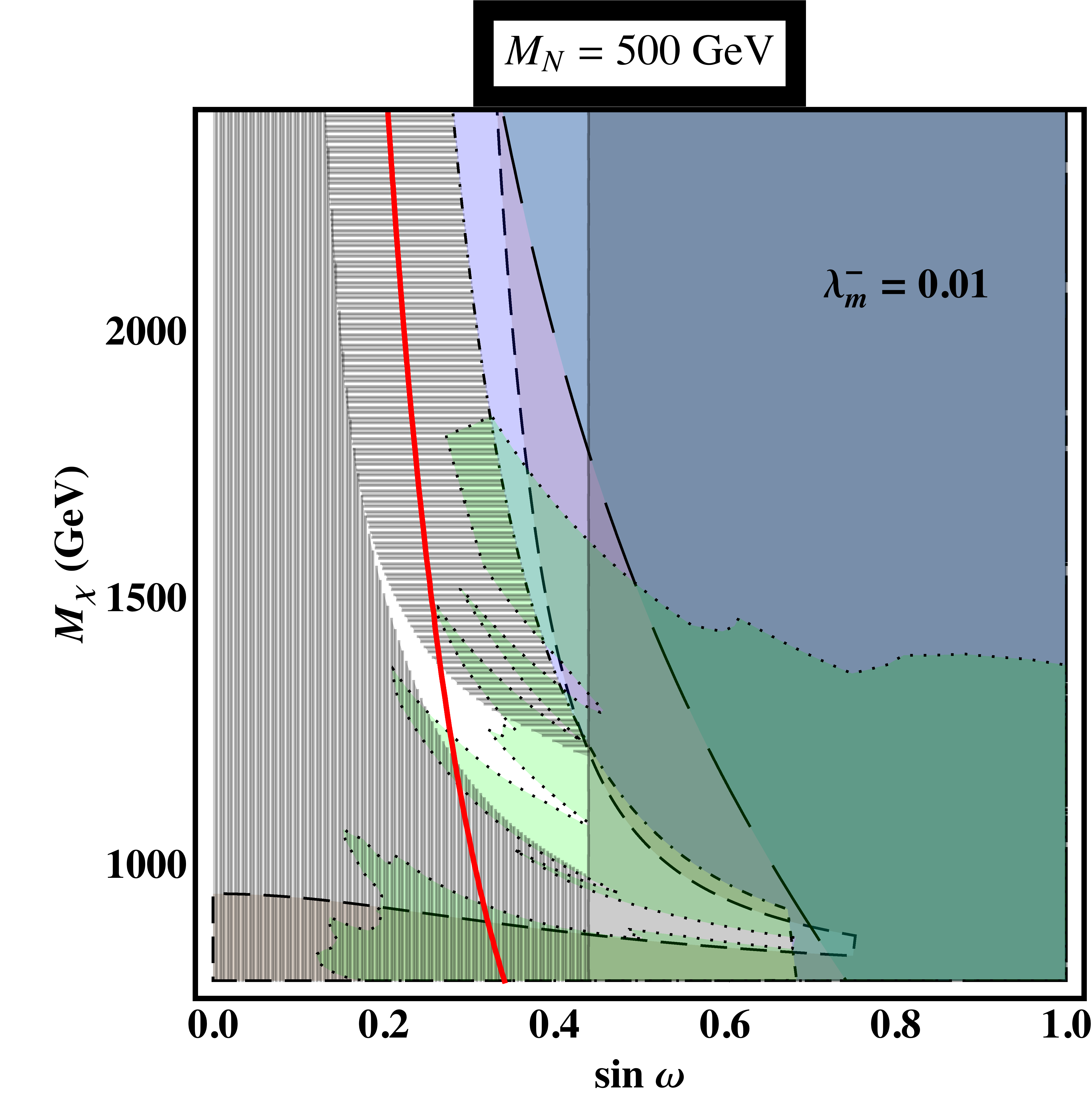}
\includegraphics[width=.44\textwidth]{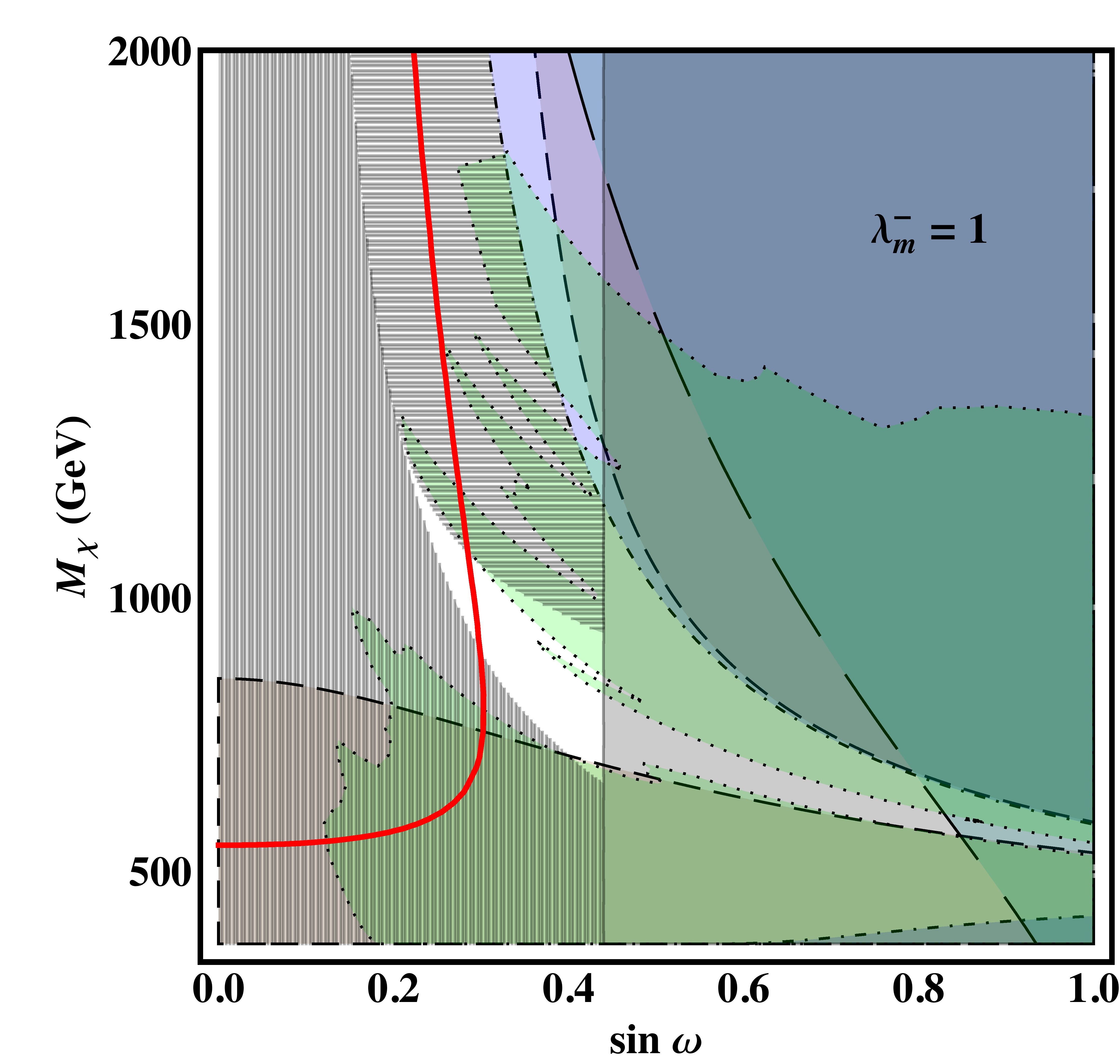}
\includegraphics[width=.44\textwidth]{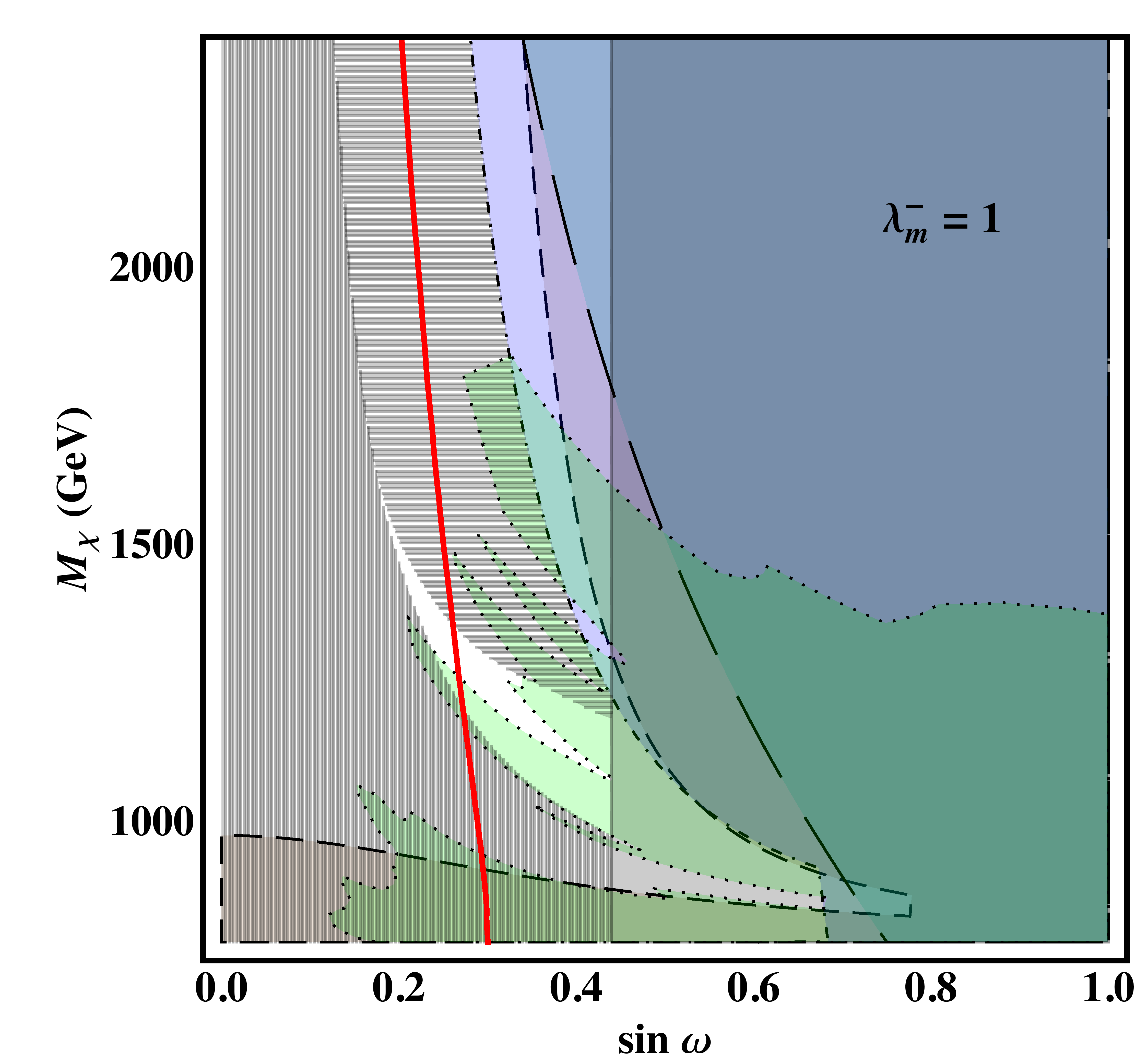}
\includegraphics[width=.44\textwidth]{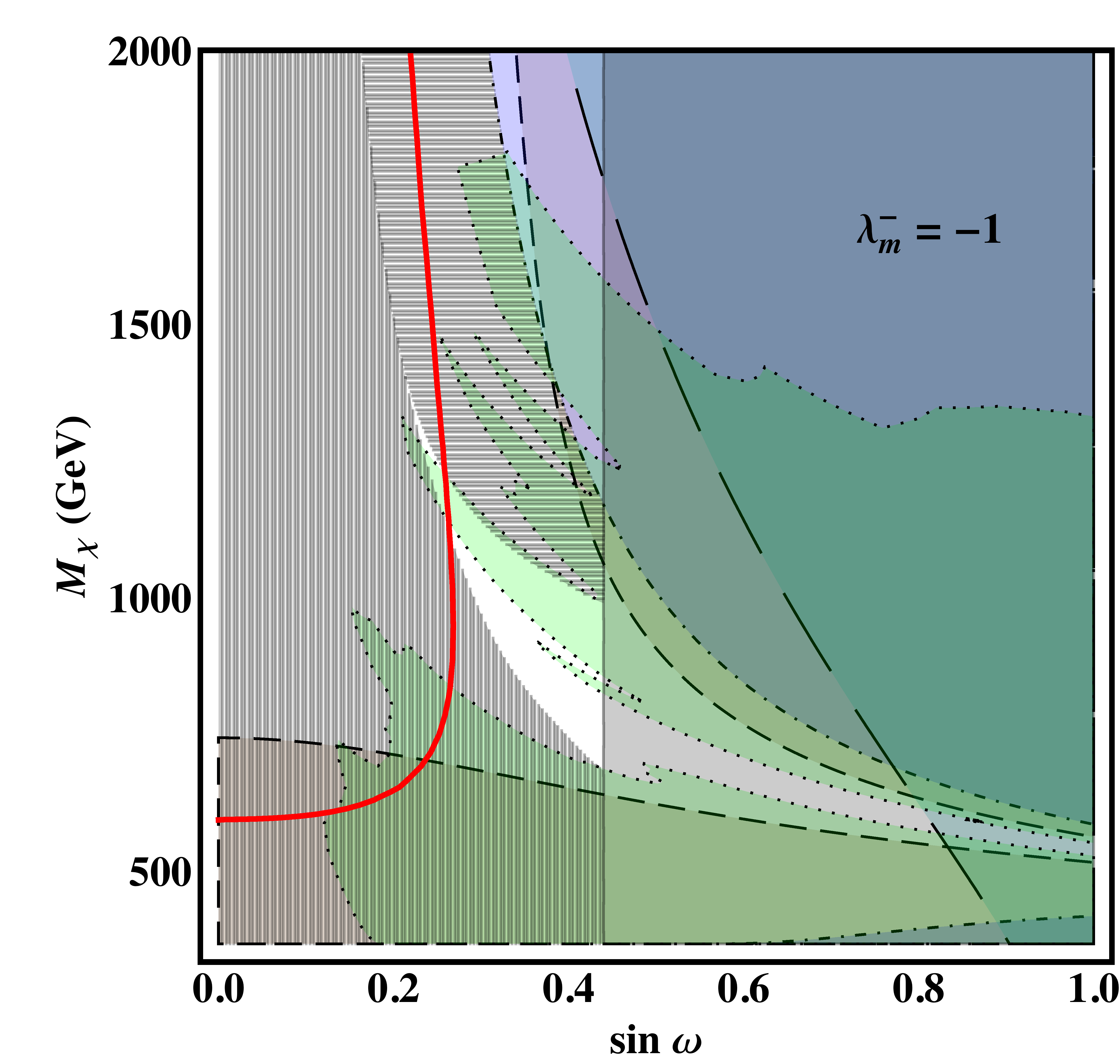}
\includegraphics[width=.44\textwidth]{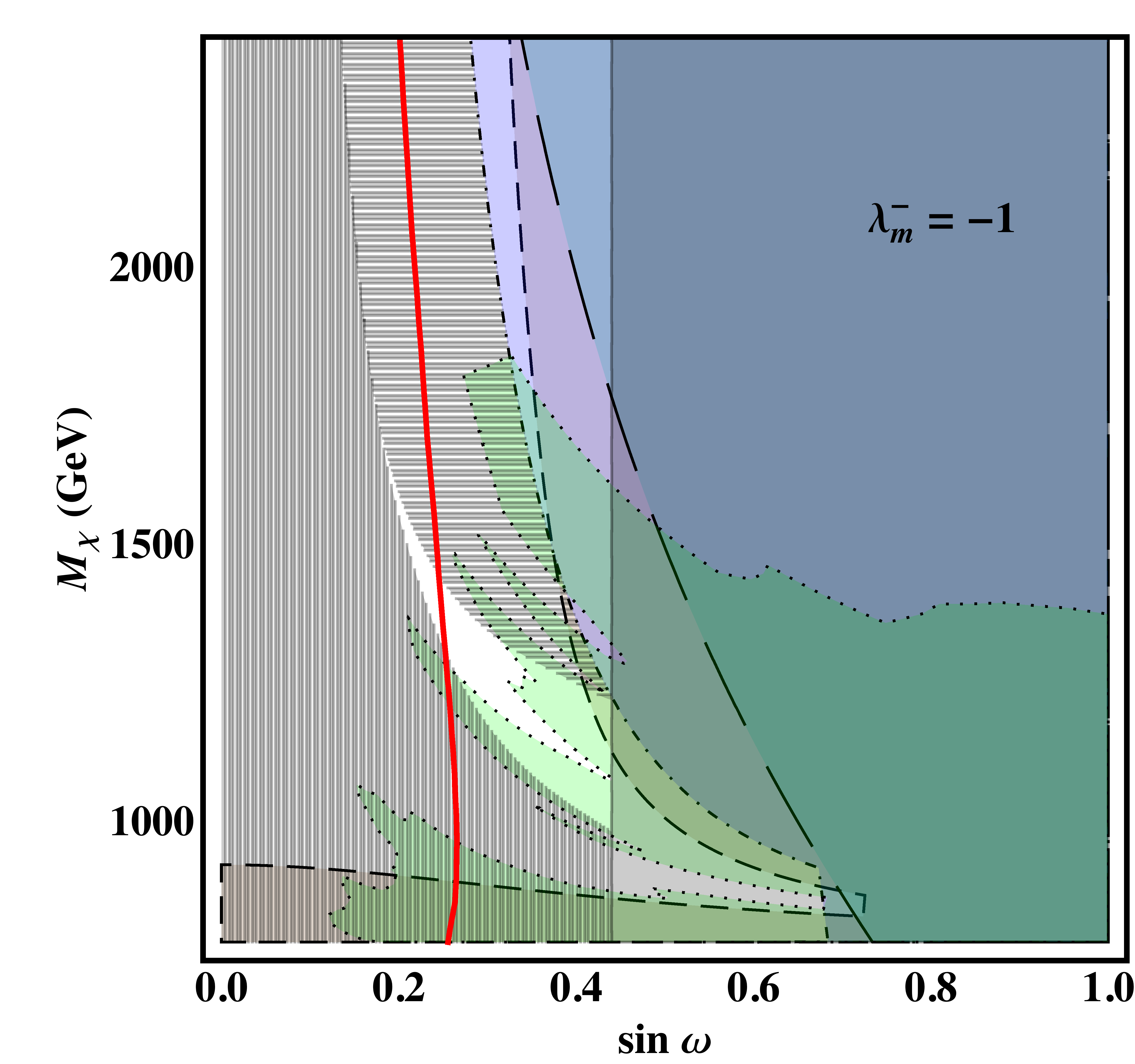}
\caption{Formal and experimental constraints in the enlarged $\sin\omega - M_{\chi}$ plane, for $\lambda_{\chi}=4\pi$ and representative values of $\lambda_{m}^{-}$ (rows) and $M_{N}$ (columns). All colored regions are excluded, with only the uncolored parts of the (center) wedge region unconstrained. The contribution of the $\chi$~pseudoscalar to the dark matter in the universe is chosen to be 3\%, $f_{\text{DM}} = 0.03$, mitigating the tension between the dark matter and strongly first-order phase transition constraints. (See the caption of Fig.~\ref{MXomSmallruledout} for the details of the plots)}
\label{MXom}
\end{figure}

Figs.~\ref{MXomSmall}~and~\ref{MXom} depict, in the enlarged $\sin \omega - M_{\chi}$~plane, the previously mentioned theoretical and experimental bounds, along with the phase transition constraints (horizontally and vertically-shaded regimes), for the benchmark values $\lambda_{\chi} = 0.1$ and $\lambda_{\chi} = 4\pi$, respectively, while imposing the (illustrative) condition $f_{\text{DM}} = 0.03$. In other words, in these figures, only 3\% of the available dark matter in the universe is attributed to the $\chi$~pseudoscalar. Comparing with the corresponding Figs.~\ref{MXomSmallruledout}~and~\ref{MXomruledout}, where $f_{\text{DM}} = 1$, one notes that the dark matter direct detection constraints are alleviated, imposing virtually no limits on the viable wedge region, and the relic abundance contour can now be accommodated within this viable region. Hence, attributing an $\mathcal O(0.01)$~fraction of the dark matter in the universe\footnote{Depending on the selected values of the input parameters, the fraction may constitute $\lesssim 10\%$ near the edge of the viable region.} to the scenario's $\chi$~pseudoscalar is compatible with the strongly first-order phase transition constraints.


To summarize, the figures illustrate that realizing a strongly first-order electroweak phase transition within the considered framework imposes considerable restrictions on the parameter space, making the scenario highly predictive. In particular, right-handed Majorana neutrino masses above several hundreds of GeV are heavily disfavored, and, depending on $M_{N}$, pseudoscalar dark matter masses are restricted within $M_{\chi} \sim 1$-2~TeV. Given the prominent role of the singlet in facilitating the realization of the strongly first-order electroweak phase transition within the scenario, the mixing angle cannot be too small, and is predicted to lie within the range $0.2 \lesssim \sin\omega \lesssim 0.4$ (the current upper bound being determined by the electroweak precision tests and the $h$ and $\sigma$~bosons' collider constraints). A dependence on the remaining input parameters $\lambda_{m}^{-}$ and  $\lambda_{\chi}$ is quite small. Moreover, the contribution of the $\chi$~pseudoscalar WIMP to the total amount of the dark matter present in the universe is confined to an $\mathcal O(0.01)$~fraction, accommodating the corresponding relic density and direct detection constraints within the viable region of the parameter space, with the nucleation temperature $T_{N} \sim100$-200~GeV.

\begin{figure}
\includegraphics[width=.43\textwidth]{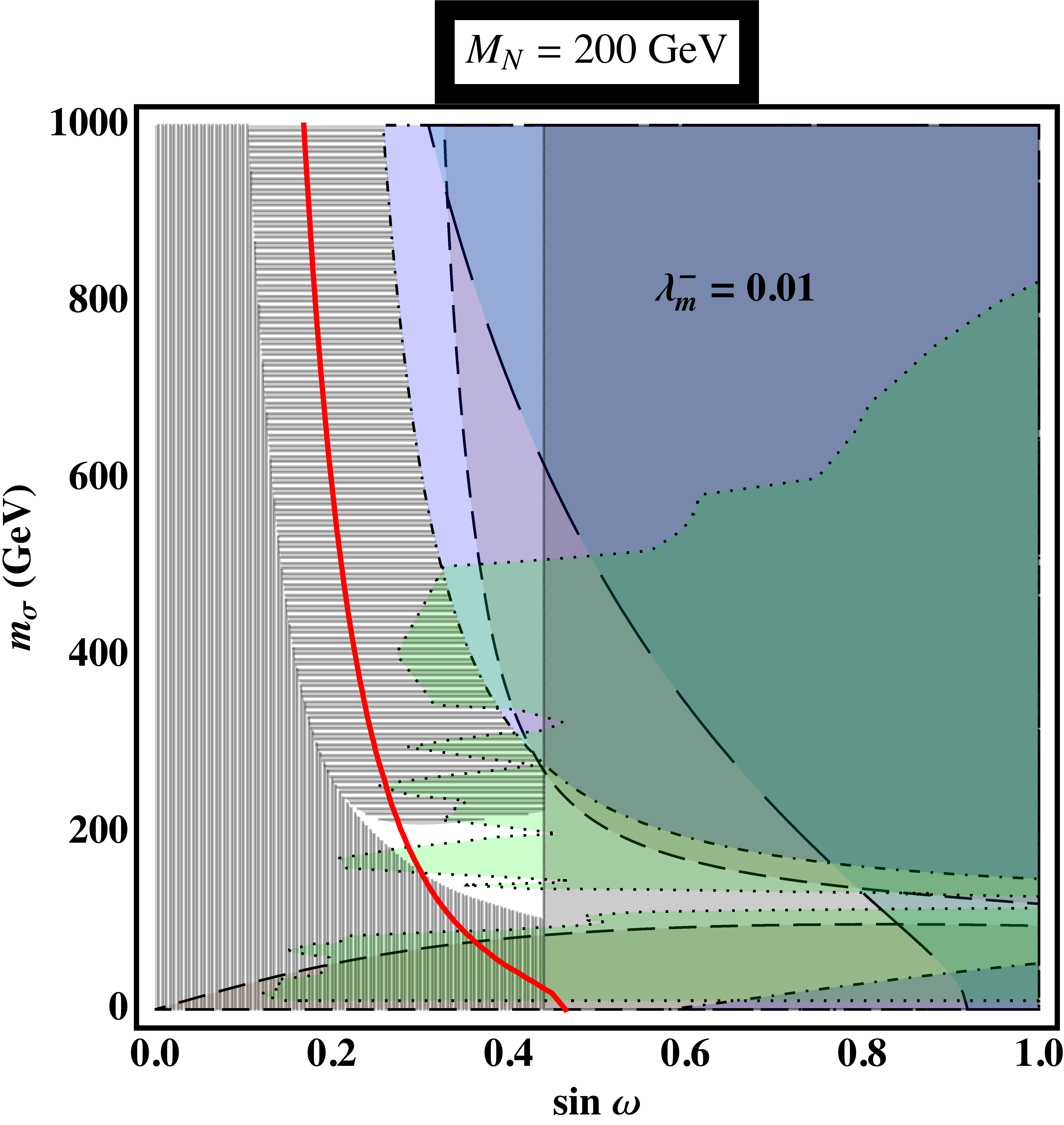}
\includegraphics[width=.43\textwidth]{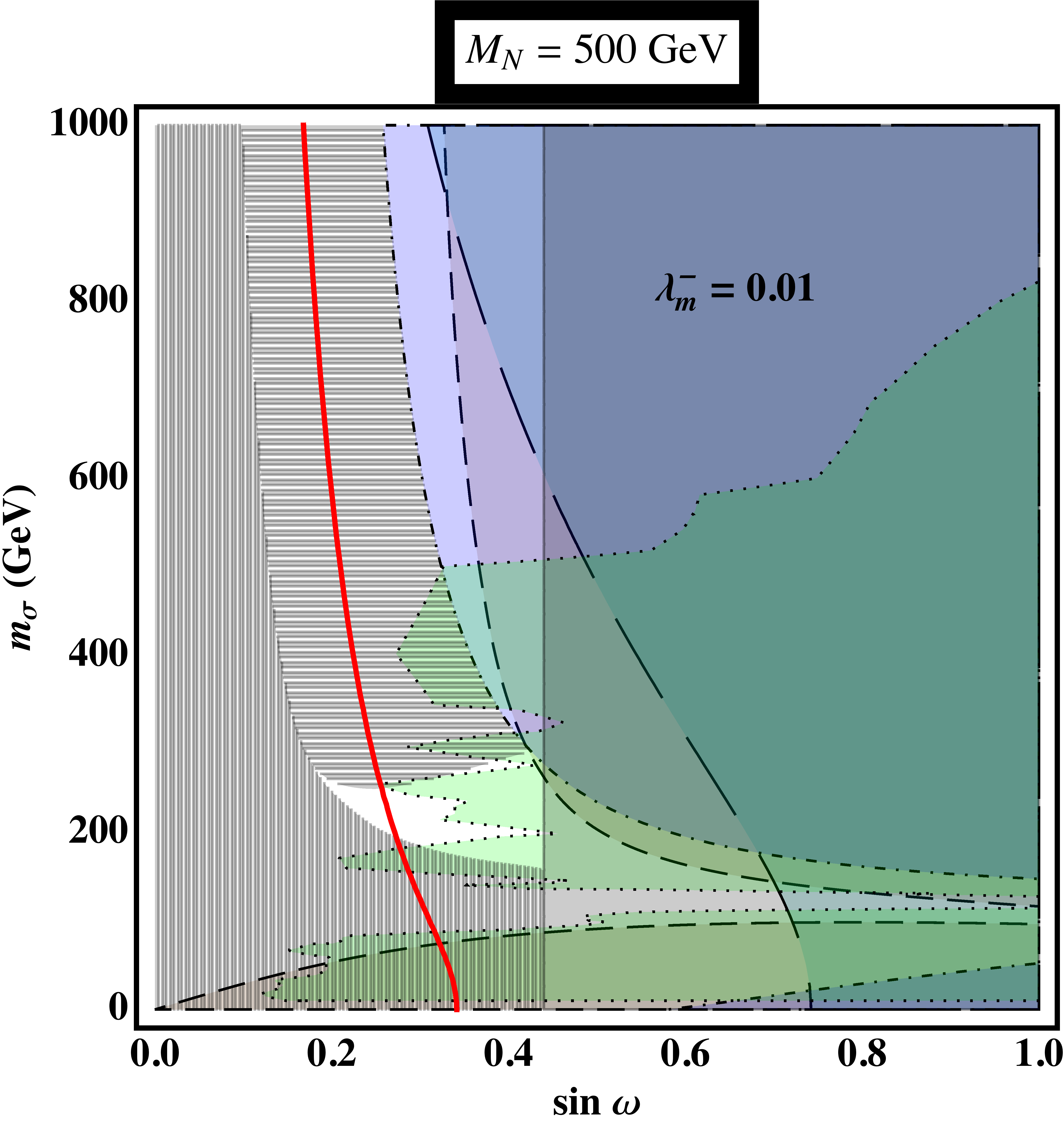}
\caption{Formal and experimental constraints in $m_{\sigma} - M_{\chi}$ plane, for $\lambda_{\chi}=0.1$ and representative values of $\lambda_{m}^{-}$ and $M_{N}$. All colored regions are excluded, with only the uncolored parts of the (center) wedge region unconstrained. The contribution of the $\chi$~pseudoscalar to the dark matter in the universe is now chosen to be 3\%, $f_{\text{DM}} = 0.03$. (See the caption of Fig.~\ref{MXomSmallruledout} for the details of the plots)}
\label{MXmsSmall}
\end{figure}

\begin{figure}
\includegraphics[width=.43\textwidth]{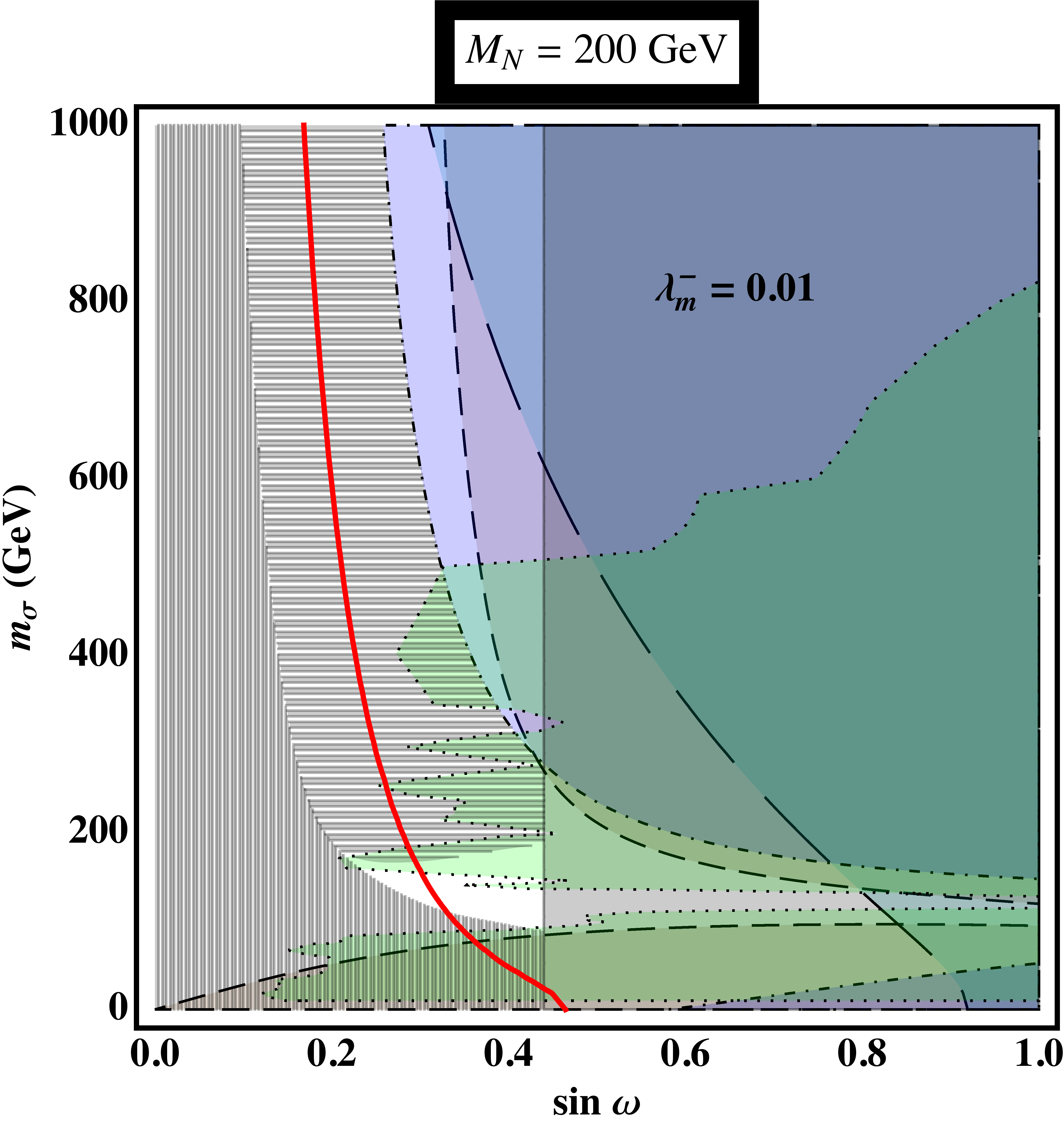}
\includegraphics[width=.43\textwidth]{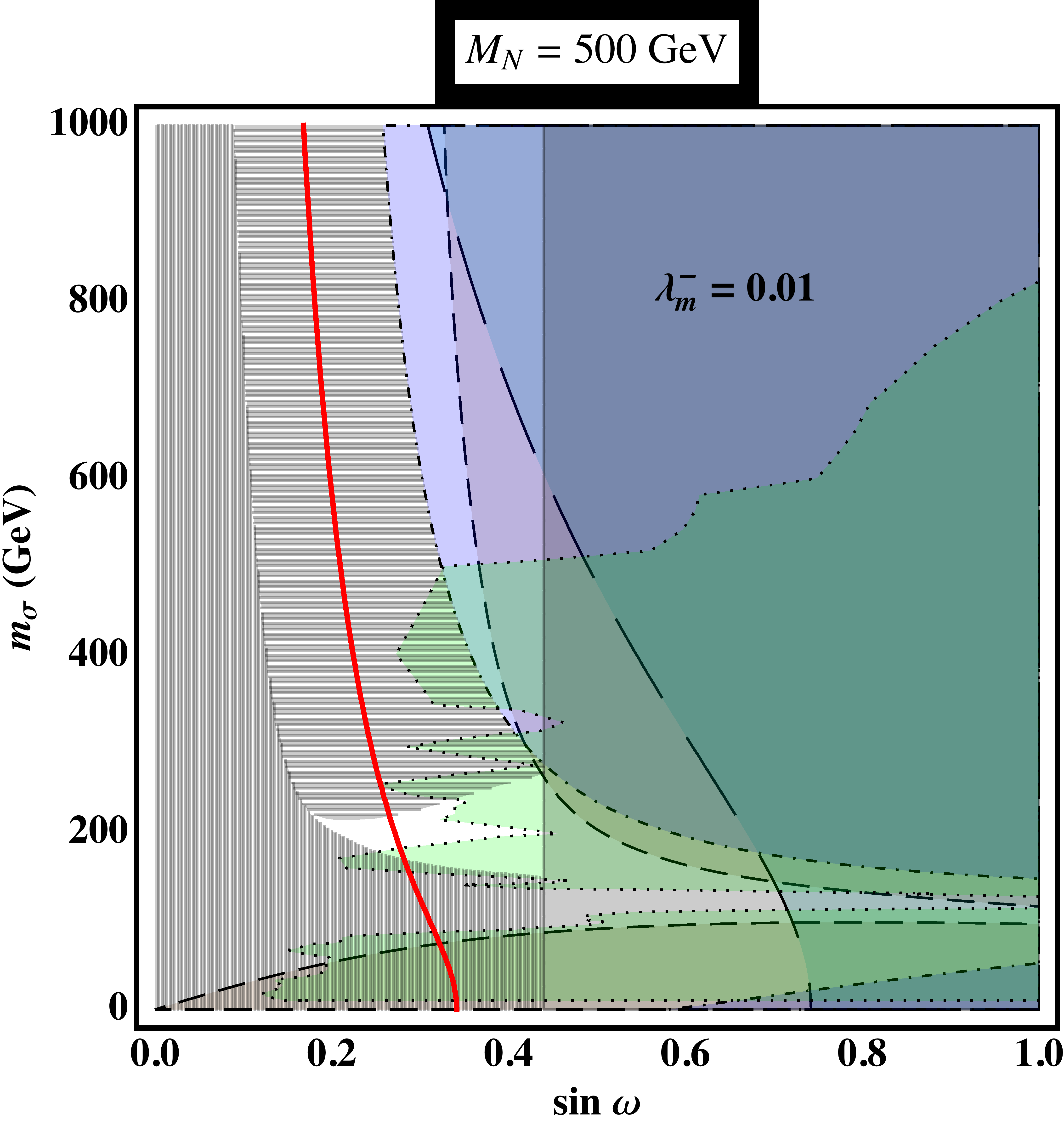}
\includegraphics[width=.43\textwidth]{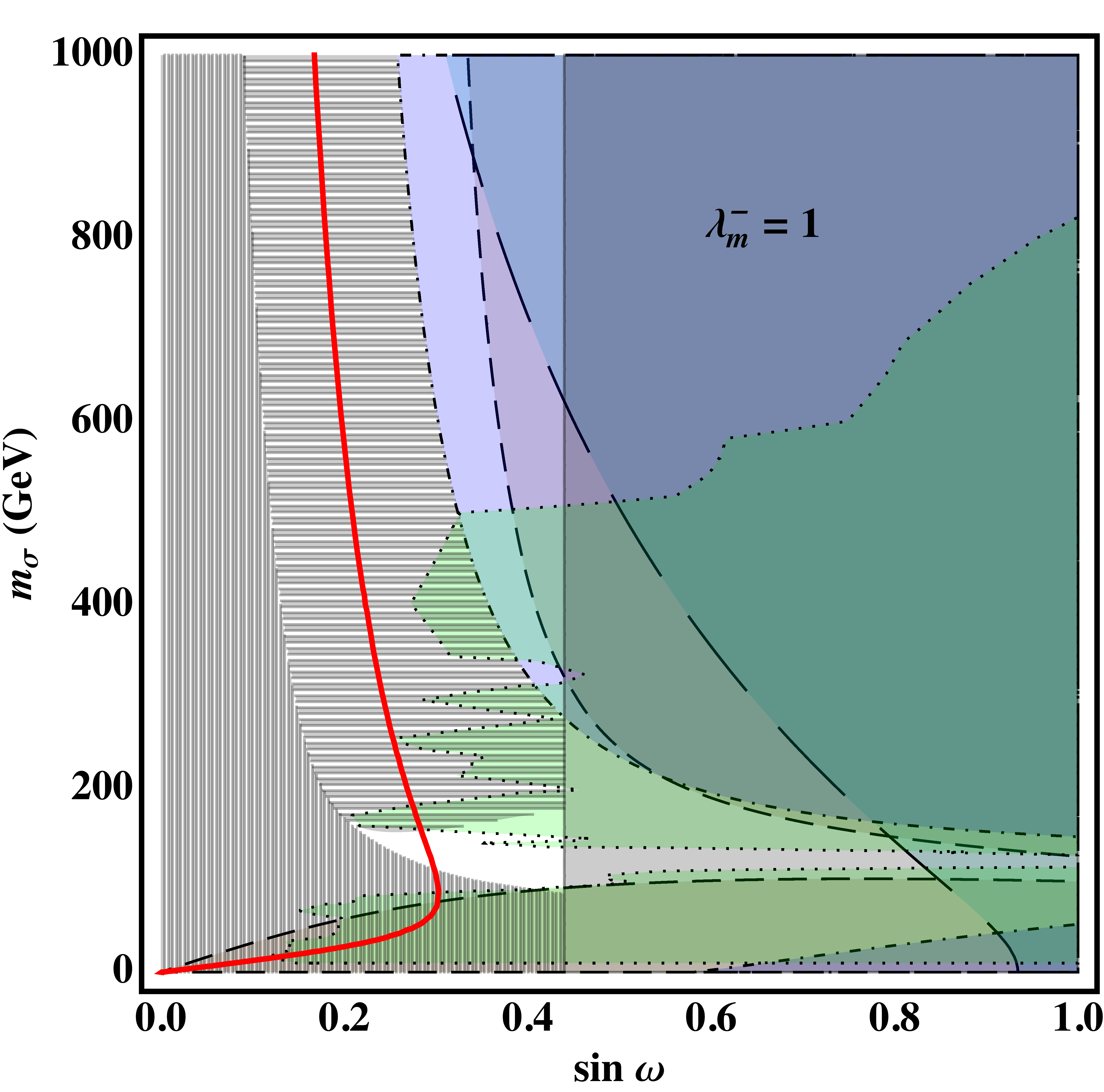}
\includegraphics[width=.43\textwidth]{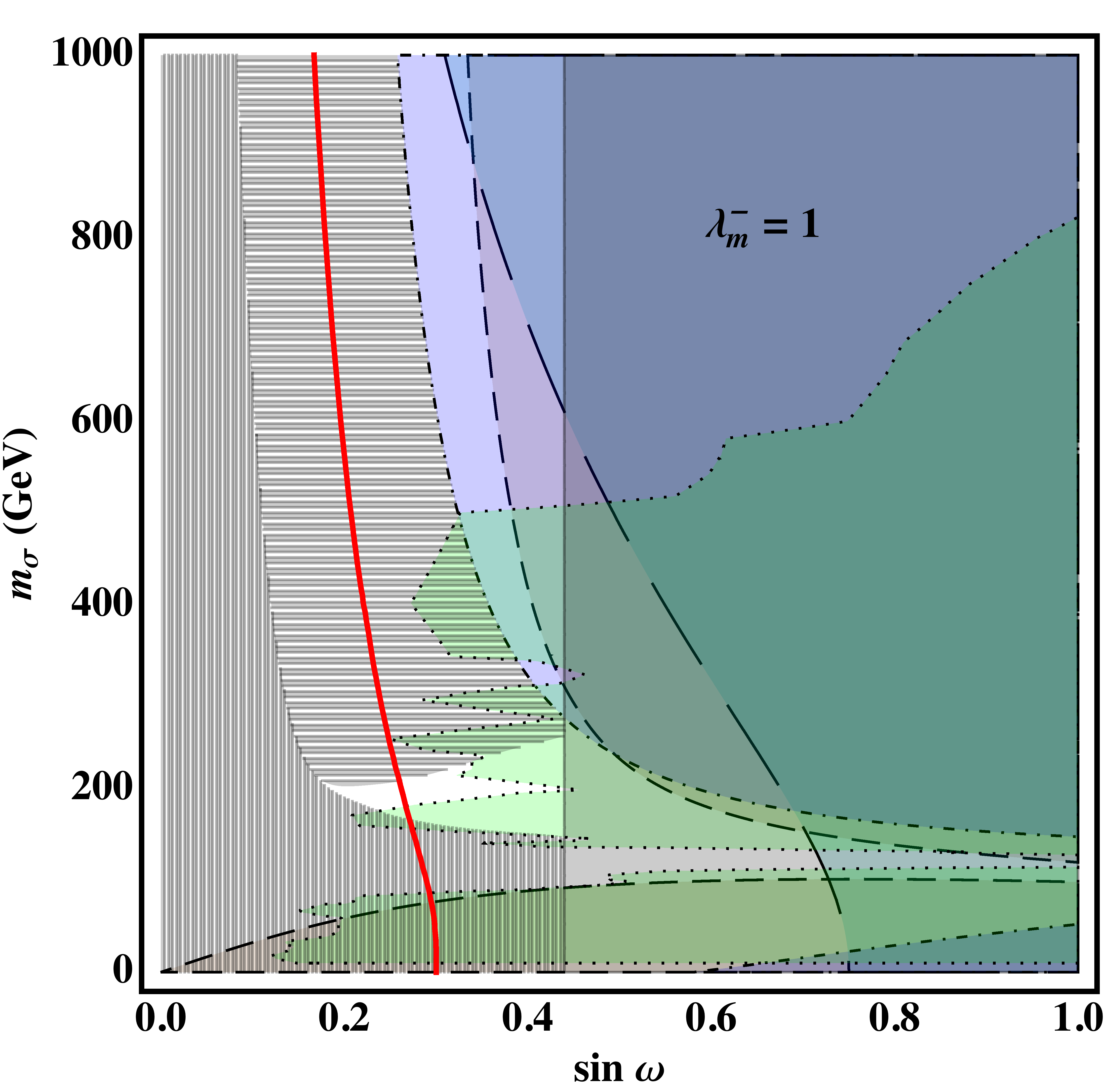}
\includegraphics[width=.43\textwidth]{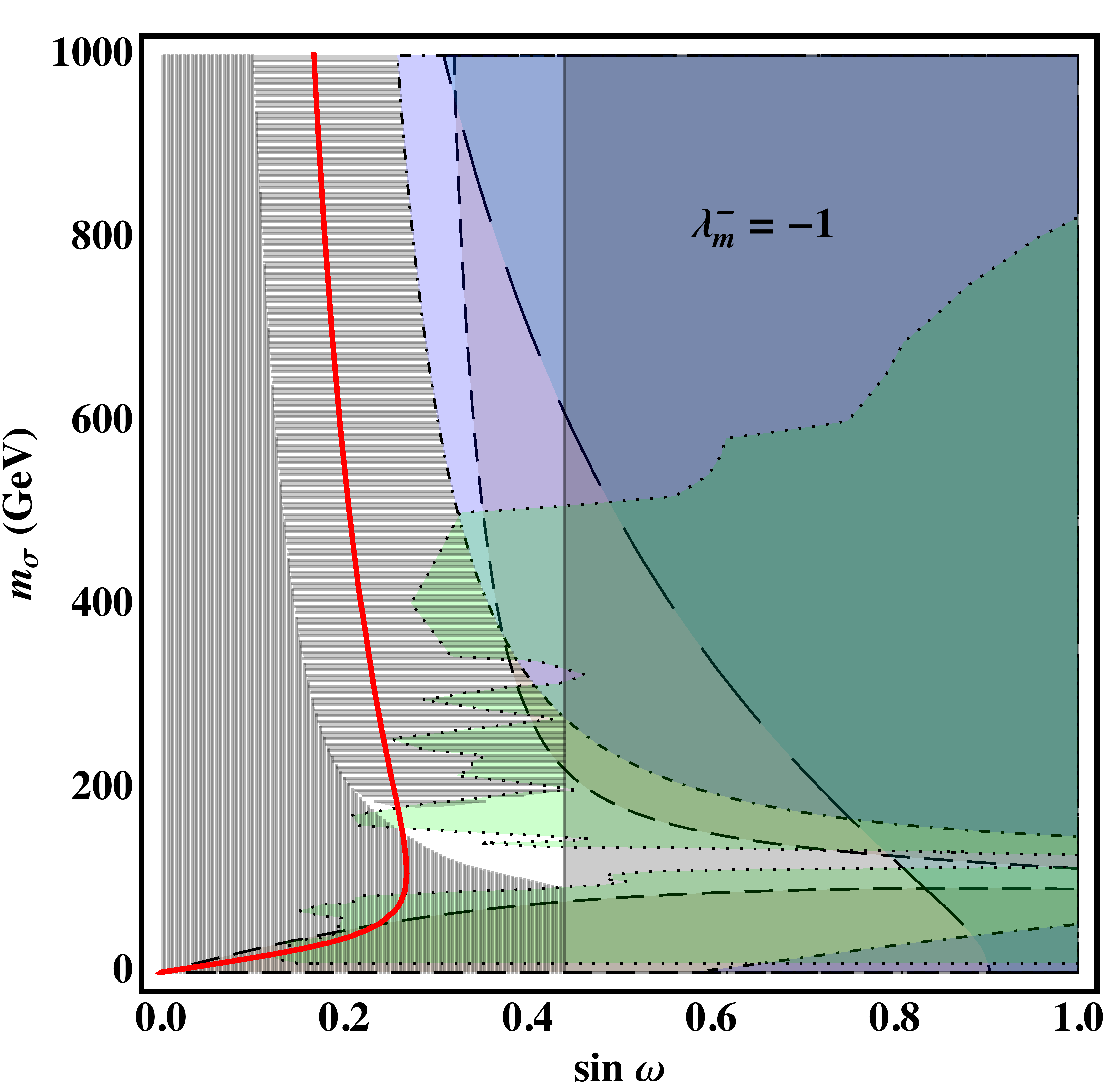}
\includegraphics[width=.43\textwidth]{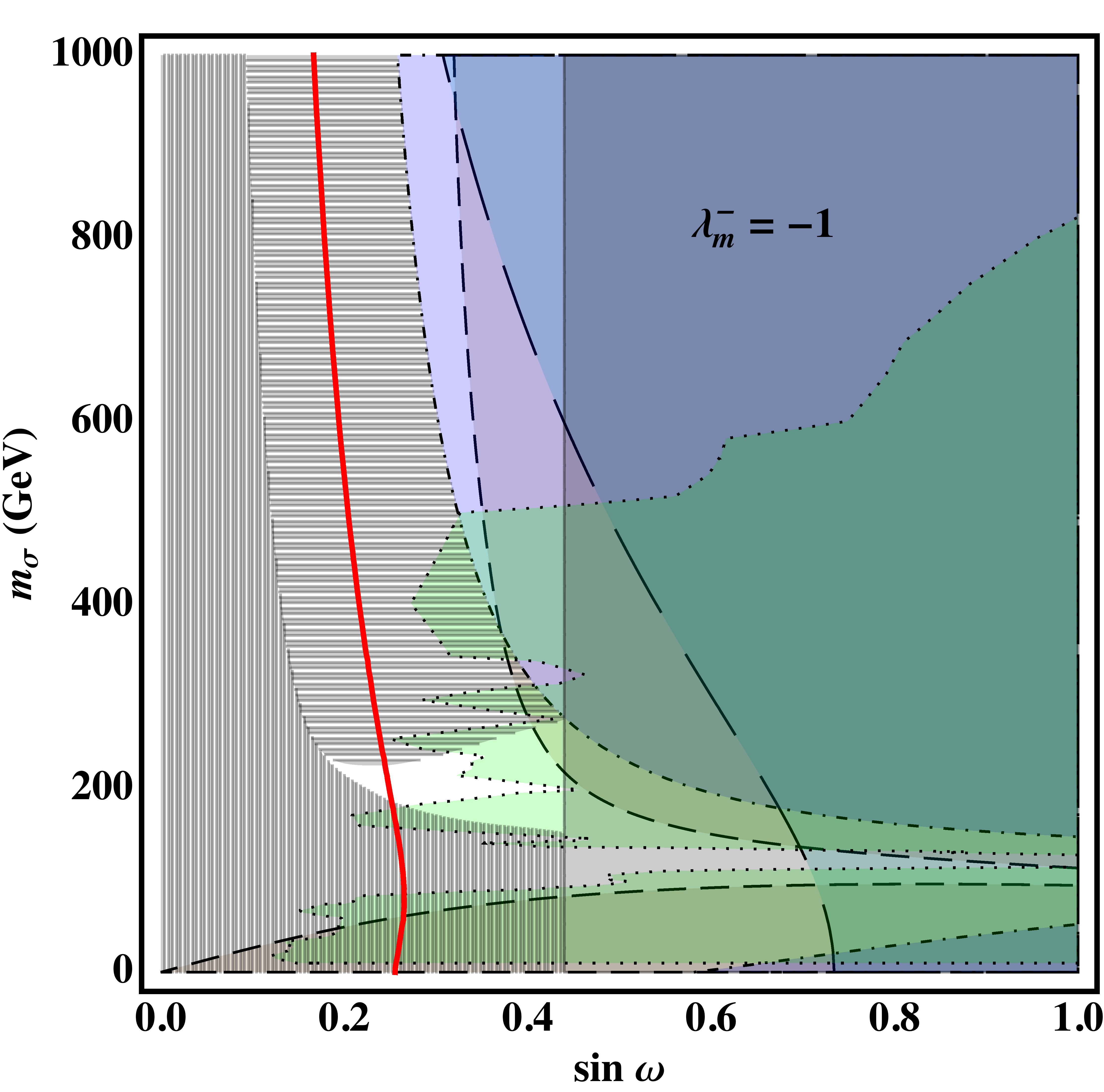}
\caption{Formal and experimental constraints in $m_{\sigma} - M_{\chi}$ plane, for $\lambda_{\chi}=4\pi$ and representative values of $\lambda_{m}^{-}$ (rows) and $M_{N}$ (columns). All colored regions are excluded, with only the uncolored parts of the (center) wedge region unconstrained. The contribution of the $\chi$~pseudoscalar to the dark matter in the universe is now chosen to be 3\%, $f_{\text{DM}} = 0.03$. (See the caption of Fig.~\ref{MXomSmallruledout} for the details of the plots)}
\label{MXms}
\end{figure}

Furthermore, It is interesting to investigate the consequences of our obtained results for the (radiatively-generated) mass of the second $CP$-even scalar degree of freedom, the $\sigma$~boson (c.f. \eqref{ms}). To this end, we plot the discussed constraints in the $\sin\omega - m_{\sigma}$~plane in Figs.~\ref{MXmsSmall}~and~\ref{MXms} (for $\lambda_{\chi} = 0.1$ and $\lambda_{\chi} = 4\pi$, respectively). As before, an illustrative $\chi$~WIMP contribution of 3\% to the dark matter content, $f_{\text{DM}} = 0.03$, has been considered. It is evident, once more, that requiring the framework to achieve a strongly first-order electroweak phase transition severely constraints its parameter space. Specifically, the mass of the $\sigma$~boson is constrained within the range $m_{\sigma} \sim 100$-300~GeV, depending on the values of the remaining input parameters, while, simultaneously, realizing an $\mathcal O(0.01)$~fraction of the dark matter in the universe.


Finally, let us briefly comment on the validity of the thin-wall approximation in the current minimal scenario. As alluded to, this approximation is valid within the regime where the thickness of the expanding bubble wall may be neglected as compared with its radius $r_{b}$. Following \cite{S3} (see also \cite{TWA}), the thickness of the bubble wall may be characterized by the expression $1/\sqrt{\frac{d^{2} V^{(0+1)}}{d\phi^{2}}}$; whereas, the radius of the expanding bubble is given by
\begin{equation}\label{rT}
r_{b} =  \frac{2}{\left | V^{(0+1)} \right |} \, S_{1}  \ ,
\end{equation}
with, $S_{1}$ defined in \eqref{S1}. Both the radius and the wall thickness of expanding bubble are, therefore, varying functions of the model's input set \eqref{inputs}. Hence, in order to examine the validity of the thin-wall approximation, one may employ the following estimation
\begin{equation}\label{TWAcond}
r_{b} \gg \tbrac{\frac{d^{2} V^{(0+1)}}{d\phi^{2}}}^{-1/2}  \ .
\end{equation}
We have verified that the condition \eqref{TWAcond} is satisfied within the entire considered range of the parameters, relevant for deriving the strongly first-order electroweak phase transition bounds. In particular, it is accurately satisfied in the vicinity of and within the viable wedge region of the parameter space. Accordingly, the employed thin-wall approximation formalism throughout our analyses is justified.

\section{Conclusion}\label{concl}

A strongly first-order electroweak phase transition is an essential ingredient for the successful implementation of an electroweak baryogenesis framework, in order to prevent the washout of the matter-antimatter asymmetry. This treatment has been devoted to examining the possibility of achieving a strongly first-order electroweak phase transition within the proposed scenario of the minimal classically scale invariant extension of the SM \cite{Farzinnia:2013pga,Farzinnia:2014xia}. In this scenario, the SM scalar content is augmented by the addition of a complex gauge-singlet state, with a potential which respects both the scale and $CP$ symmetry. The latter symmetry prevents the pseudoscalar degree of freedom from decaying, rendering it a suitable WIMP dark matter candidate. In addition, inclusion of the right-handed Majorana neutrinos allows for properly accounting for the non-zero masses of the SM neutrinos, via the seesaw mechanism.

In order to explore the nature and the strength of the phase transition, the full finite-temperature effective potential of the model at one-loop has been constructed, including the contributions from the resummed thermal bosonic daisy loops. Despite the absence of a barrier in the zero-temperature one-loop effective potential, it was demonstrated that the finite-temperature effects induce, at this order, a barrier between the symmetric and the broken phase vacua, and, thereby, give rise to a first-order electroweak phase transition. We required the first-order phase transition to be sufficiently strong at the onset of bubble nucleation, in order to prevent the washout of a potential matter-antimatter asymmetry, and demonstrated that such a requirement  imposed formidable constraints on the model's input parameter space.

In particular, it was found that the constraints from a strongly first-order electroweak phase transition were severely in tension with the previously analyzed dark matter relic density and direct detection bounds \cite{Farzinnia:2014xia}, where it was presumed that the dark matter in the universe was dominantly composed of the scenario's pseudoscalar. Similar results have been obtained within the context of the ($Z_{2}$-symmetric) non-scale invariant scenarios \cite{DM&EWPT}, and we have demonstrated the validity of this conclusion in the considered minimal classically scale symmetric model. Nevertheless, relaxing the assumption regarding the single-component nature of the dark matter, it was shown that an $\mathcal O(0.01)$ fraction of the dark matter in the universe can be accommodated by the scenario's pseudoscalar, while, simultaneously realizing a strongly first-order electroweak phase transition within the considered minimal framework. A summary of the obtained results are displayed in Figs.\ref{MXomSmall}-\ref{MXms}, where the derived bounds are exhibited in combined exclusion plots, covering relevant representative values of the input parameters, and identifying the (small) viable regions of the parameter space (corresponding to nucleation temperatures $T_{N} \sim100$-200~GeV).

We conclude that the considered minimal classically scale invariant scenario is capable of realizing a strongly first-order electroweak phase transition, which imposes powerful constraints on its free parameter space, and considerably boosts its predictivity: an $\mathcal O(0.01)$ fraction of the dark matter in the universe may be composed of the scenario's pseudoscalar, the right-handed Majorana neutrino masses heavier than several hundreds of GeV are disfavored, pseudoscalar dark matter mass is confined to $\sim 1$-2~TeV, and the mass of the second $CP$-even scalar lies within the range $\sim 100$-300~GeV, depending on the choices of the input parameters. Moreover, the mixing angle between the $CP$-even components of the SM doublet and the complex singlet is predicted to be $0.2 \lesssim \sin\omega \lesssim 0.4$. The current upper bound on the mixing angle is determined by the LHC measurements of the properties of the $h$~Higgs, the collider searches for the $\sigma$~boson, as well as the data from the electroweak precision tests (c.f. Fig.~\ref{LHCexp}). Hence, many of these predictions may be further probed by the next LHC run.

\section*{Acknowledgment}

We are grateful to R. Sekhar Chivukula for valuable comments on the early versions of the manuscript, and to Thomas Konstandin for useful correspondence. A.F. thanks Kenji Kadota for interesting discussions. During the completion of this work, A.F. was in part supported by the Tsinghua Outstanding Postdoctoral Fellowship and by the NSF of China (under grants 11275101, 11135003). J.R. was in part supported by National NSF of China (under grants 11275101, 11135003) and National Basic Research Program (under grant 2010CB833000).

\appendix*

\section{Feynman Rules}\label{FR}

\begin{figure}
\includegraphics[width=.85\textwidth]{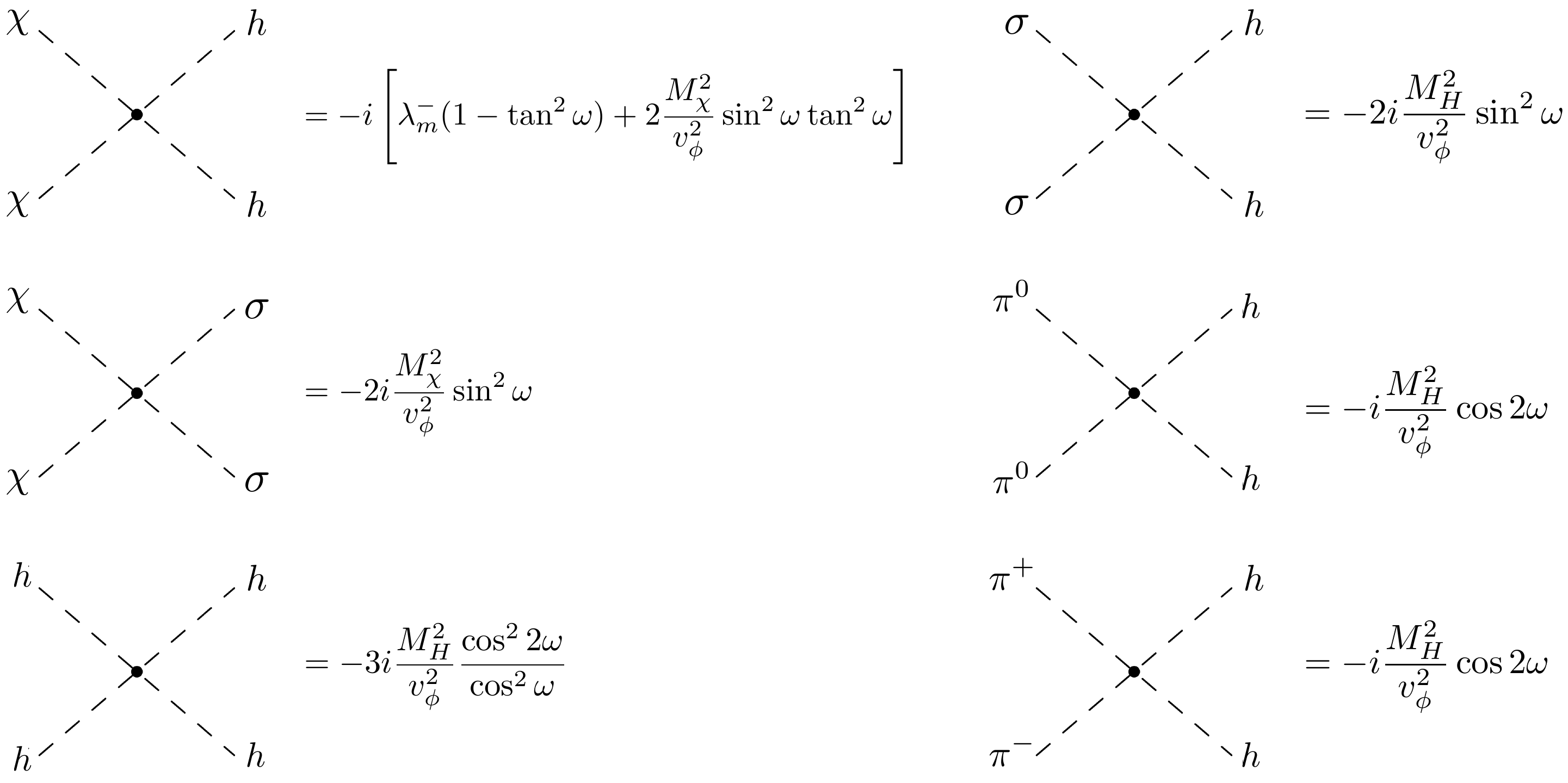}
\caption{Feynman rules for the relevant quartic couplings, $i\lambda_{ijkl}$.}
\label{FR4}
\end{figure}

Fig.~\ref{FR4} exhibits the quartic couplings' Feynman rules, relevant for computing the thermal masses \eqref{mtherm} which enter the finite-temperature one-loop effective potential.

\end{document}